\documentclass[11pt,preprint]{aastex}

\usepackage{multicol}
\usepackage{nicefrac}

\shorttitle{The Role of Environment in Compact Groups}
\shortauthors{Walker et al.}

\begin{document}

\title{Examining the Role of Environment in a Comprehensive Sample of Compact Groups}

\author{Lisa May Walker and Kelsey E. Johnson}
\affil{Department of Astronomy, University of Virginia,
    Charlottesville, VA 22904}
\author{Sarah C. Gallagher}
\affil{Department of Physics and Astronomy, University of Western Ontario,
    London, ON N6A 3K7 Canada}
\author{Jane C. Charlton}
\affil{Department of Astronomy and Astrophysics, Pennsylvania State University,
    University Park, PA 16802}
\author{Ann E. Hornschemeier}
\affil{Laboratory for X-Ray Astrophysics, NASA Goddard Space Flight Center,
    Greenbelt, MD 20771}
\and
\author{John E. Hibbard}
\affil{National Radio Astronomy Observatory,
    Charlottesville, VA 22903}

\begin{abstract}
Compact groups, with their high number densities, small velocity dispersions, and an interstellar medium that has not been fully processed, provide a local analog to conditions of galaxy interactions in the earlier universe. The frequent and prolonged gravitational encounters that occur in compact groups affect the evolution of the constituent galaxies in a myriad of ways, for example gas processing and star formation. Recently, a statistically significant ``gap'' has been discovered mid-infrared (MIR: 3.6-8$\mu$m) IRAC colorspace of compact group galaxies. This gap is not seen in field samples and is a new example of how the compact group environment may affect the evolution of member galaxies. In order to investigate the origin and nature of this gap, we have compiled a larger sample of 37 compact groups in addition to the original 12 groups studied by \citet{johnson07} (yielding 174 individual galaxies with reliable MIR photometry). We find that a statistically significant deficit of galaxies in this gap region of IRAC colorspace is persistant in the full sample, lending support to the hypothesis that the compact group environment inhibits moderate SSFRs. Using this expanded sample, we have more fully characterized the distribution of galaxies in this colorspace and quantified the low density region more fully with respect to MIR bluer and MIR redder colors. We note a curvature in the colorspace distribution, which is fully consistent with increasing dust temperature as the activity in a galaxy increases. This full sample of 49 compact groups allows us to subdivide the data according to physical properties of the groups. An analysis of these subsamples indicates that neither projected physical diameter nor density show a trend in colorspace within the values represented by this sample. We hypothesize that the apparent lack of a trend is due to the relatively small range of properties in this sample, whose groups have already been pre-selected to be compact and dense. Thus, the relative influence of stochastic effects (such as the particular distribution and amount of star formation in individual galaxies) becomes dominant. We analyze spectral energy distributions of member galaxies as a function of their location in colorspace and find that galaxies in different regions of MIR colorspace contain dust with varying temperatures and/or PAH emission.
\end{abstract}

\keywords{galaxies: evolution --- galaxies: interactions --- galaxies: clusters --- galaxies: statistics --- infrared: galaxies}

\section{INTRODUCTION}\label{intro}
The majority of galaxies reside in poor groups of galaxies \citep{mulchaey00}. With number densities similar to those seen in the centers of rich clusters, compact groups constitute a high-density sub-category of poor groups \citep{hickson82}. This environment is thought to provide a local analog to conditions in the early universe when the average galaxy density was higher and interactions were more common and prolonged.

Despite their similar densities, compact groups exhibit marked differences from the centers of clusters. This is especially true with regards to the interstellar medium (ISM), a crucial component in the process of galaxy evolution. For instance, the centers of clusters tend to have no neutral gas, while compact groups typically contain at least some neutral gas, though they are known to be \ion{H}{1} deficient as a class \citep{verdes01}. These \ion{H}{1} deficiencies in compact groups are likely related to the ISM behavior in the outskirts of clusters, where the degree of \ion{H}{1} deficiency is correlated with the radial distance of the galaxy to the cluster's center \citep{giovanelli85}. In addition, both compact groups and clusters show a correlation between \ion{H}{1} deficiency and X-ray detectability \citep{giovanelli85,verdes01}, suggesting that the neutral gas undergoes a phase change via ionization and heating.

A key difference between compact groups and galaxy clusters is that galaxies in the cores of clusters typically have negligible star formation \citep{balogh98,linden10}, while some compact groups can host moderate or even intense star formation \citep{iglesiasparamo99,gallagher10}. However, IRAS observations of compact groups indicate normal levels of thermal infrared emission, suggesting that the star formation rates (SFRs) in these groups are not generally enhanced \citep{allam95,verdes98}. In fact, \citet{rosa07} studied the stellar populations of elliptical galaxies in compact groups and find evidence of a mechanism that quenches star formation. This is again similar to the outer regions of clusters, which exhibit suppression of SFRs out to three virial radii \citep{lewis02}. Gas processing is also different in cluster cores and compact groups, where ram-pressure stripping does not appear to dominate gas processing \citep{rasmussen08}. In addition, \citet{cortese06} have studied a compact group falling into a cluster, and found that the galaxies have undergone preprocessing due to the compact group environment. Despite their high densities, it appears as though the compact group environment has more in common with the outer regions of clusters than with cluster cores.

Recently, \citet{walker10} examined the distribution of galaxies from 12 compact groups in IRAC (3.6-8.0 $\mu$m) colorspace and found evidence for a statistically significant gap between the location of galaxies with colors consistent with normal stellar populations and galaxies with colors indicative of star formation activity \citep[previously identified by][]{johnson07, gallagher08}. In contrast, no gap is seen in the colorspace distribution of the core of the Coma cluster, a control sample of field galaxies, or a sample of interacting galaxies. While this is not surprising, the varying distributions in IRAC colorspace indicate that these environments have had a significant impact on the evolution of their galaxies. However, a similar (though less pronounced) gap was observed in the outer regions of the Coma cluster. This is evidence that the gap is unique to regions of enhanced galaxy density (relative to the field) in which the neutral gas has not been fully processed.

The discovery of the gap in IRAC colorspace naturally leads to the question of the origin of the gap, which has been intractable due to the small size of the previous sample. Initial results comparing the IRAC gap with specific SFRs \citep[SSFRs;][]{tzanavaris10} of the same sample of groups show that the gap in IRAC colorspace corresponds to a gap in SSFRs \citep{walker10}, suggesting a shared underlying cause. In order to better investigate the region of colorspace identified as the gap, we need to increase the number of observed compact group galaxies, so as to understand the characteristics of any systems that may be found in this region as well as characterize any correlation with physical properties along the sequence in colorspace. In order to further populate IRAC colorspace, in this paper we present the colors of galaxies for a large sample of Hickson Compact Groups \citep[HCG;][]{hickson82} and Redshift Survey Compact Groups \citep[RSCG;][]{barton96} at redshifts of $z < 0.035$ with a full suite of IRAC data available from the {\it Spitzer Space Telescope} archive, yielding 49 compact groups including 179 galaxies.

In this paper, we adopt the following terms: the ``original sample'' is the 12 HCGs studied in \citet{johnson07} and subsequent papers. The ``new sample'' consists of the 21 HCGs and 16 RSCGs not previously studied by this group. The ``full'' or ``expanded sample'' is the combined original and new samples.

\section{DATA}
\subsection{Samples}
\subsubsection{Compact Groups}\label{cgsample}
Because the goal of this project is to understand galaxy evolution in the dense environment of compact groups, we began with the HCG and RSCG catalogs. These catalogs were compiled using different selection criteria, thus the groups may have slightly different characteristics; however the RSCG criteria were chosen to create a catalog with properties similar to the HCG catalog. The HCG catalog was compiled by \citet{hickson82} using photometric plates from the Palomar Sky Survey. The criteria to be classified as a compact group were N $\ge$ 4 (four members within 3 magnitudes of the brightest galaxy), $\theta_N \ge 3\theta_G$ (no other galaxies located within a radius 3 times the radius containing the nuclei of the members), and $\mu_G < 26.0$ (the surface brightness must be high). The RSCG catalog, compiled by \citet{barton96}, utilized magnitude limited redshift surveys to locate compact groups and determine their members. The criteria to be a member of a compact group were to have a radial velocity difference of $\Delta V \le 1000\;{\rm km\,s^{-1}}$ and galaxy separation of $D \le 50\;{\rm kpc}$ between neighbors. Galaxy groups meeting these criteria were included in the RSCG sample. There was no isolation criteria for the RSCGs, and as a result many are embedded in larger structures. There are several groups that appear in both the HCG and RSCG catalog; naturally these groups have only been included once in the full sample. The RSCGs include groups with three or more members. The original criteria for HCGs was four or more members, but some have been trimmed to three because one foreground or background galaxy was found to be projected onto the group \citep{hickson92}. Including the RSCGs in the sample will allow future analysis of the effect of varying group properties and environments on the member galaxies.

For this study, the HCG and RSCG catalogs were sub-selected to only include compact groups at a low enough redshift ($z < 0.035$) such that the polycyclic aromatic hydrocarbon (PAH) features do not shift out of their rest-frame bands, which would make an analysis of their distribution in colorspace infeasible. We searched the {\it Spitzer} archive for groups meeting this criterion, and included groups for which all four channels of IRAC data are available. This yielded 21 HCGs (in addition to the original sample of 12 HCGs) and 16 RSCGs, bringing the total number of compact groups in the full sample to 49 (see Table \ref{cgtable}), and increasing the number of galaxies from 42 to 179. Of these 179 galaxies, 5 are saturated in one or more IRAC bands (HCG 16d, HCG56d, HCG 90d, HCG 92b, and RSCG 4b), and are not included in the fits or statistical analysis presented in this paper. Five of the 16 RSCGs (21, 44, 66, 67, and 68) are known to be embedded in larger structures.
We have undertaken our analysis both with and without these groups, and in most cases, the results do not change.

\subsubsection{Comparison Samples}\label{compsamples}
As in \citet{walker10}, we compare our compact group sample with samples of galaxies in other environments. These samples are our approximation to a ``field'' sample - LVL+SINGS \citep{dale09,dale07}, interacting galaxies \citep{smith07}, and two samples from the Coma cluster -- the center and the infall region \citep{jenkins07}. See \citet{walker10} for a discussion of these samples.

\begin{figure}
  \plotone{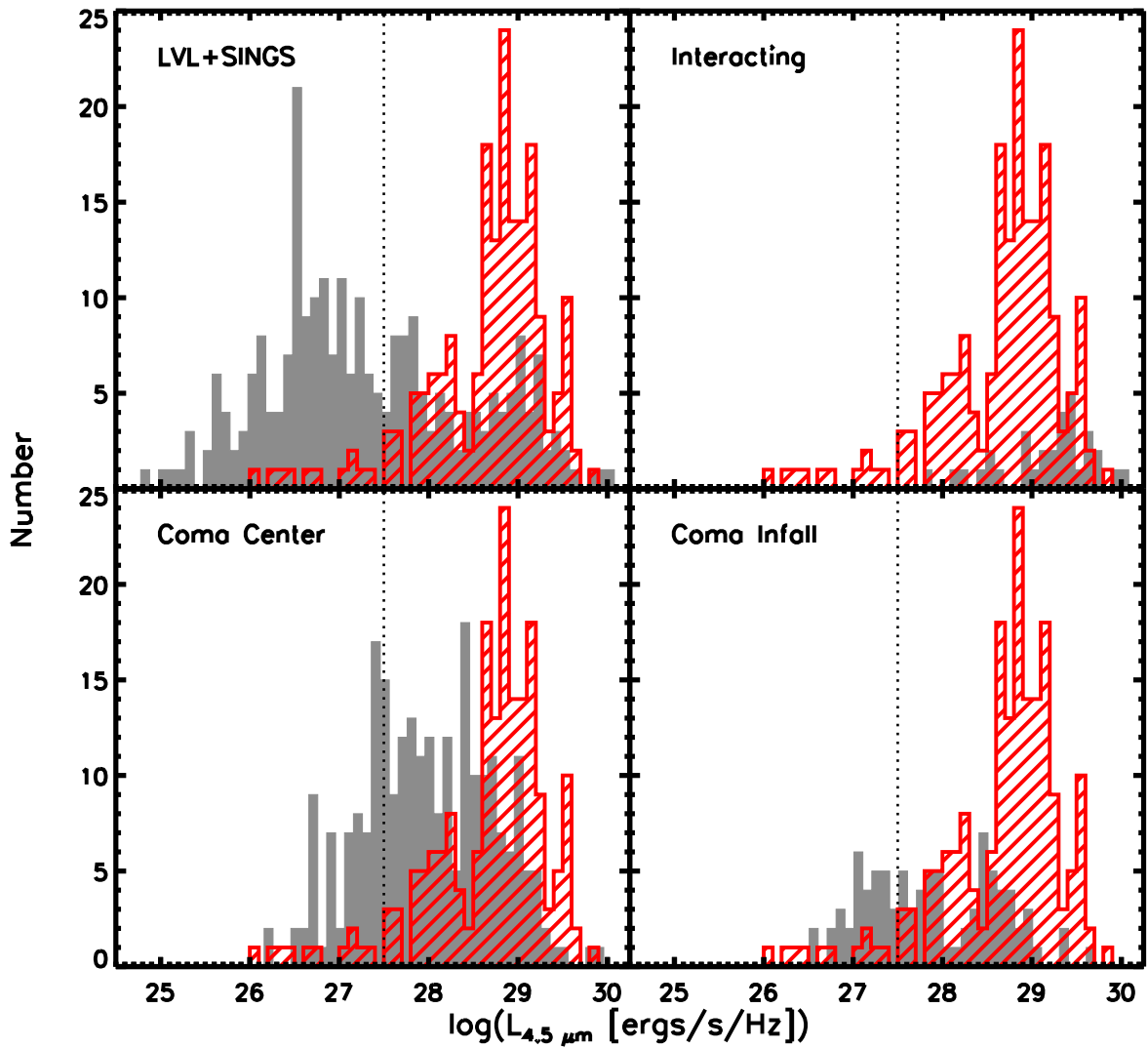}
  \caption{Histograms of $L_{4.5}$ for the compact group sample (striped) overlaid on the comparison samples (solid). The dotted vertical line indicates the luminosity cut.\label{lumhist}}
\end{figure}

In order to compare samples of similar characteristics, we apply a luminosity cut to our samples, illustrated in Figure \ref{lumhist}. We selected the lowest luminosity bin with more than two compact group galaxies as the minimum luminosity for all of our samples. Thus our luminosity cut is $\log{\left(L_{4.5}\;\left[\rm{erg/s/Hz}\right]\right)} = 27.5$. Figure \ref{morphhist} shows the Hubble types for our compact group sample against LVL+SINGS. As can be seen in this plot, the LVL+SINGS sample is dominated by late-type galaxies (which is to be expected for the field), while the compact group sample shows a bimodal distribution of morphological types, with a large number of early-type galaxies.

\subsection{Observations}
As this study utilizes archival data, the observational setups are heterogeneous. In addition to the {\it Spitzer} data, we also utilize data from the Two Micron All Sky Survey \citep{skrutskie06} to plot IR spectral energy distributions (SEDs) to shorter $\lambda$.

\subsection{Photometry}
The IRAC images were first convolved to the 8 $\mu$m PSF in order to ensure a common pixel scale and resolution, then photometry was performed using SURPHOT \citep{reines08}. The program determines apertures by finding a specified contour level in a reference image, then measuring the flux within the same aperture for each image. The apertures were determined using a combined image of all four IRAC bands (weighted by $\lambda^{-1}$), using a contour level of $1 - 2 \sigma$. Several background annuli were used, with inner radii ranging from $2-2.5\times$ the radius of the aperture and outer radii ranging from $2.5-3\times$ the radius of the aperture, using both the mode and resistant mean. Uncertainties were calculated from the standard deviation of the fluxes measured using different backgrounds. Extended source aperture corrections were made using the algorithms of T. Jarrett\footnote{\url{http://web.ipac.caltech.edu/staff/jarrett/irac/calibration/}}, and average corrections were 4\%, 2\%, 9\%, and 14\% at 3.6, 4.5, 5.8, and 8.0 $\mu$m, respectively. There are slight differences from \citet{johnson07}, likely due to variations in background measurements, though the new data are still consistent with the previous measurements.

\begin{figure}
  \plotone{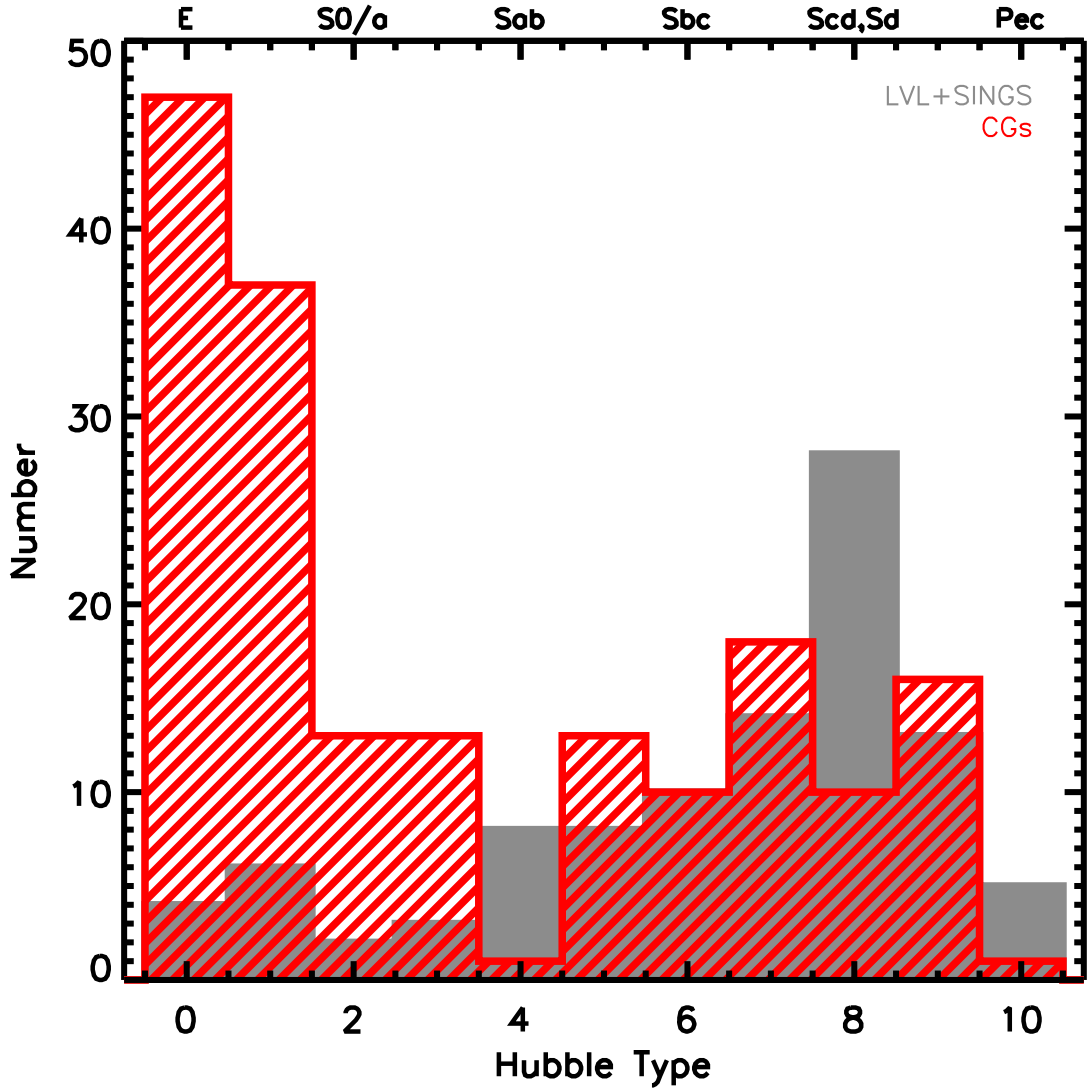}
  \caption{Distribution of morphological types \citep[following][]{haynes84} for the compact group galaxies (striped) and LVL+SINGS galaxies (solid). As expected for a field sample, the LVL+SINGS galaxies are dominated by late-types, while the CG sample has a bimodal distribution.\label{morphhist}}
\end{figure}
\input{obsinfo.tex}

\section{COLORSPACE}
\subsection{Full Sample}\label{fullcolor}
\begin{figure}
  \plotone{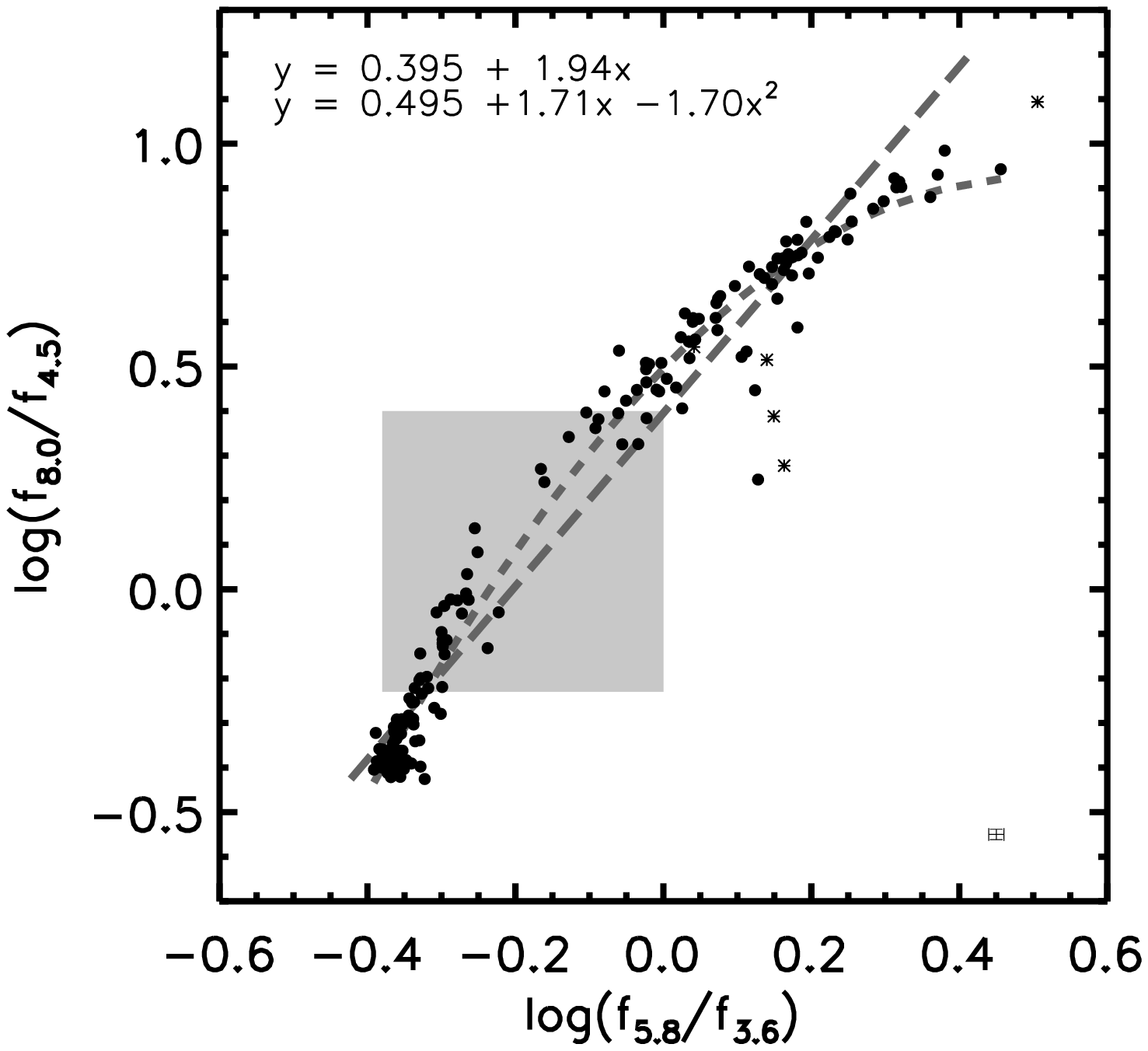}
  \caption{The colorspace distribution of the 179 compact group galaxies comprising the full sample; the shaded box highlights the location of the gap in the original sample \citep{walker10}. The asterisks indicate galaxies that are mildly saturated in one or more IRAC channels, these 5 galaxies are not included in further analysis. We performed two different fits to the data, shown above.\label{colorspace}}
\end{figure}

The IRAC colors of the full sample are presented in Figure \ref{colorspace}, which shows that an underdensity of galaxies is persistent, though it has become more akin to a canyon than a gap. As the sample is now comprised of 179 galaxies, there are more galaxies with colors that place them within the canyon, though there is still a dearth of galaxies in this region relative to the number of galaxies with MIR-blue and MIR-red colors. The plot in Figure \ref{colorspace} shows the data with a linear fit, however a new result seen in this full sample of 49 compact groups is the {\it curvature} of the distribution in colorspace, which will be discussed in \S\ref{sedsection}. To quantify this curvature, we fit a quadratic to the data, also shown in Figure \ref{colorspace}. Thus, we can undertake our analysis with respect to the line or the curve, both fit to the data.

\begin{figure}
  \plottwo{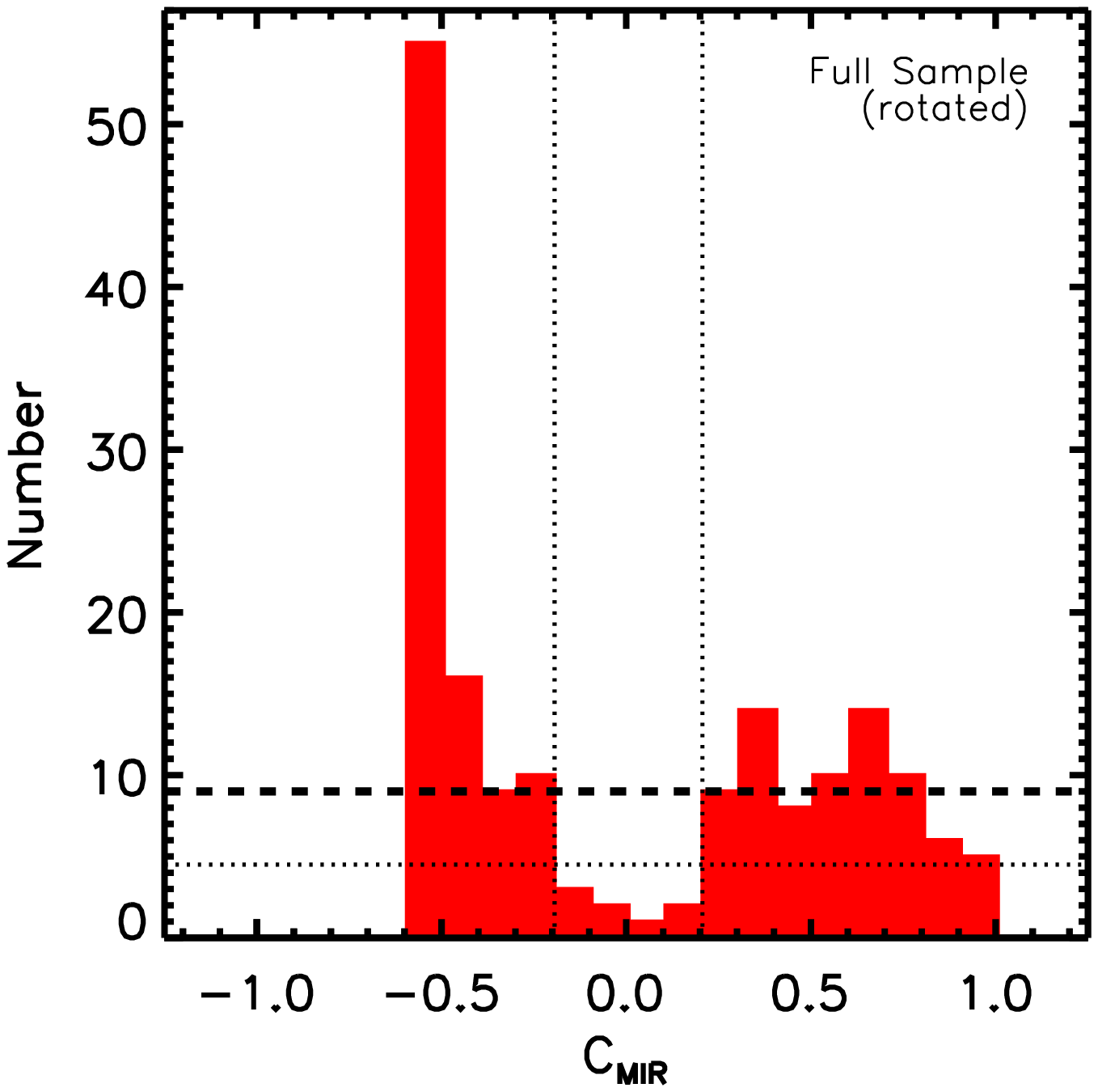}{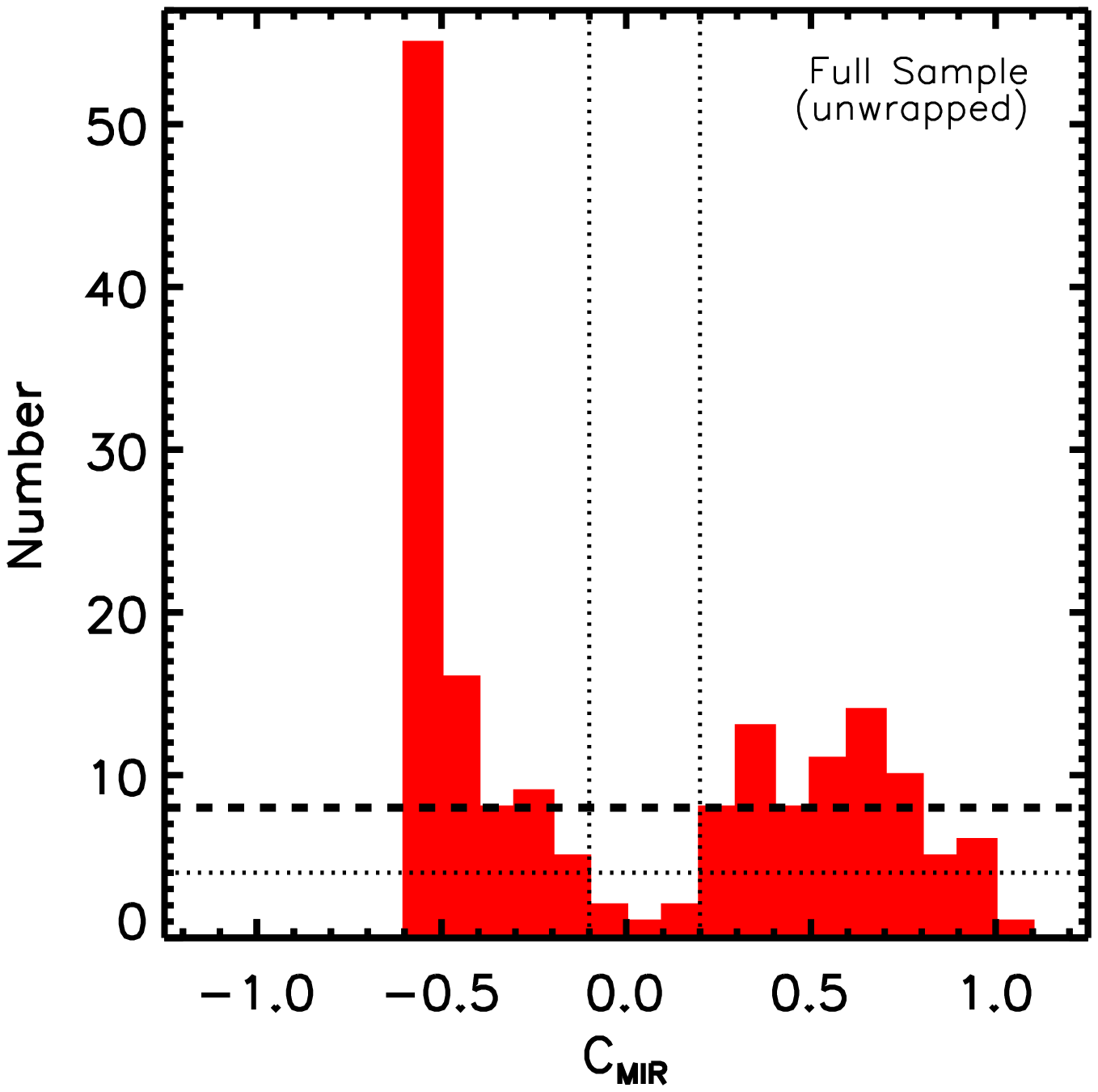}
  \caption{Histograms of the {\it left}: rotated colorspace distribution and {\it right}: unwrapped colorspace distribution as described in \S\ref{fullcolor}. In both plots, the dashed line indicates the median, the horizontal dotted line is half that value. The vertical dotted lines indicate the boundaries of the canyon, defined to be where the distribution is less than half the median value with an MIR color $<1.0$. In the rotated sample, 5\% of the galaxies fall in the canyon, while for the unwrapped sample, 3\% of galaxies fall in the canyon.\label{colhist}}
\end{figure}

For the statistical analysis, we considered the MIR colors in two ways: {\it rotated} so that the line shown in Figure \ref{colorspace} became the axis, and {\it unwrapped} so that the curve shown in Figure \ref{colorspace} became the axis. For both methods, the sample was shifted to have a mean value of zero. We have redefined the gap as the canyon using both sets of colors, as illustrated in Figure \ref{colhist}. In both plots, we define the canyon to be where the histogram is less than half its median value and $\rm{C_{MIR}}$ is $< 1.0$. The bounds of the canyon are indicated by the vertical dotted lines. The blue side of the canyon changes depending on whether we define it using the rotated or unwrapped color - the canyon determined from the unwrapped data is slightly smaller. However, the results of our statistical analysis do not change.

\input{KStest.tex}
\begin{figure}
  \plottwo{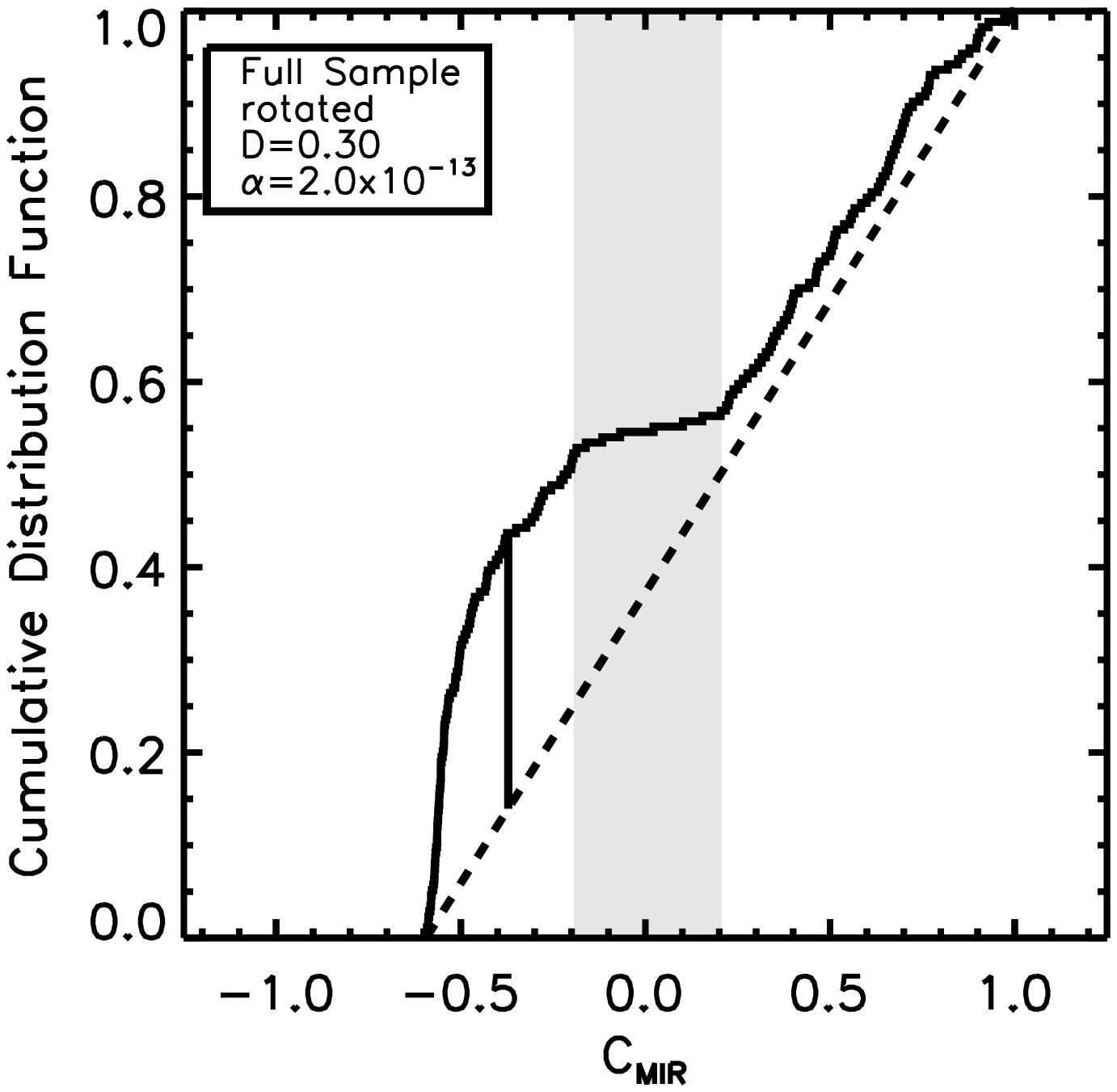}{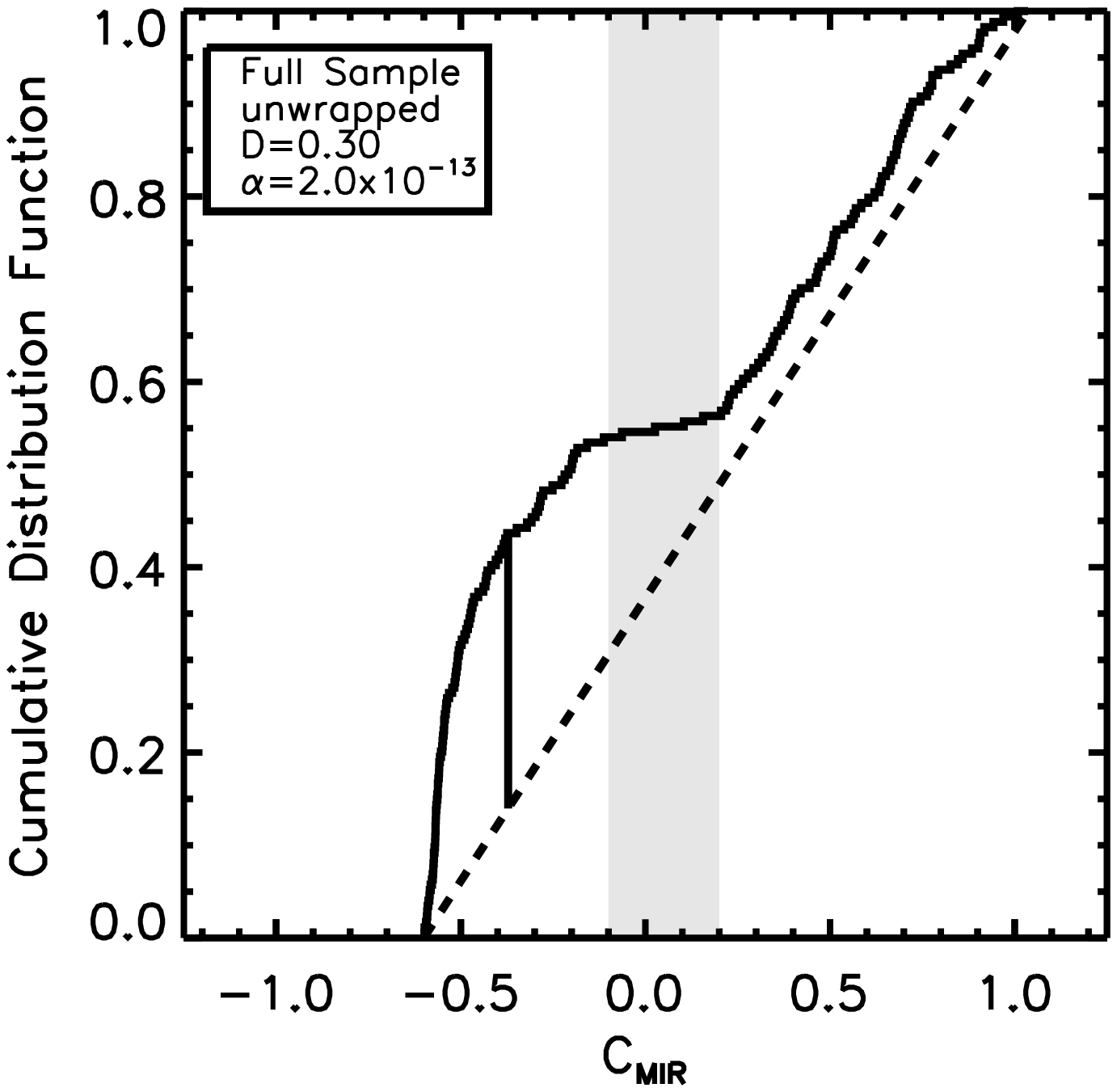}
  \caption{KS test for the {\it left}: rotated colorspace distribution and {\it right}: unwrapped colorspace distribution of the full sample against a model of uniform distribution. The maximum deviation $D$ of the sample from the model is indicated by the vertical line. This large value of $D$ means a low probability that the compact groups are drawn from a uniform distribution, $2\times10^{-11}$\%.\label{expks}}
\end{figure}

\begin{figure}
  \plottwo{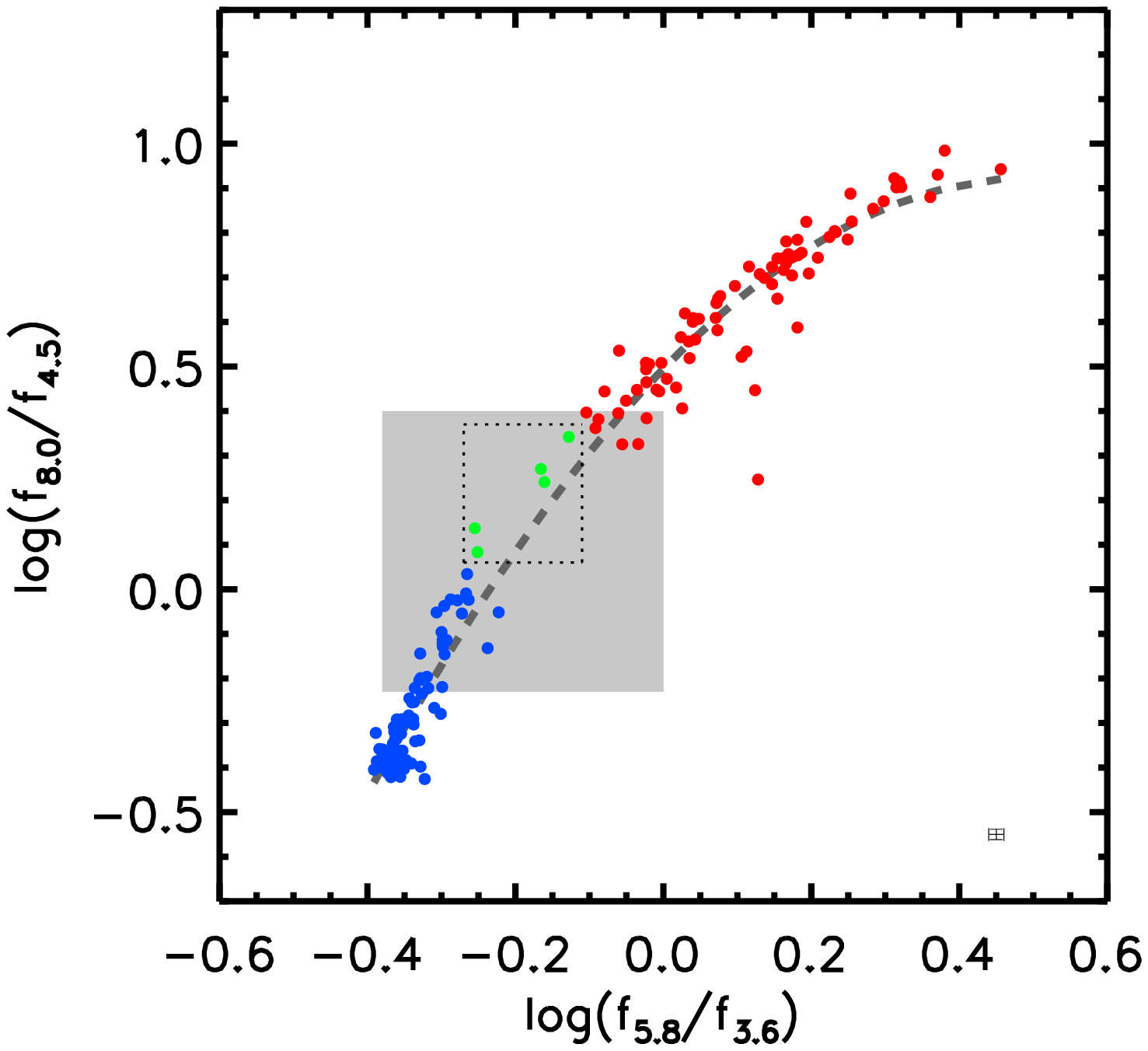}{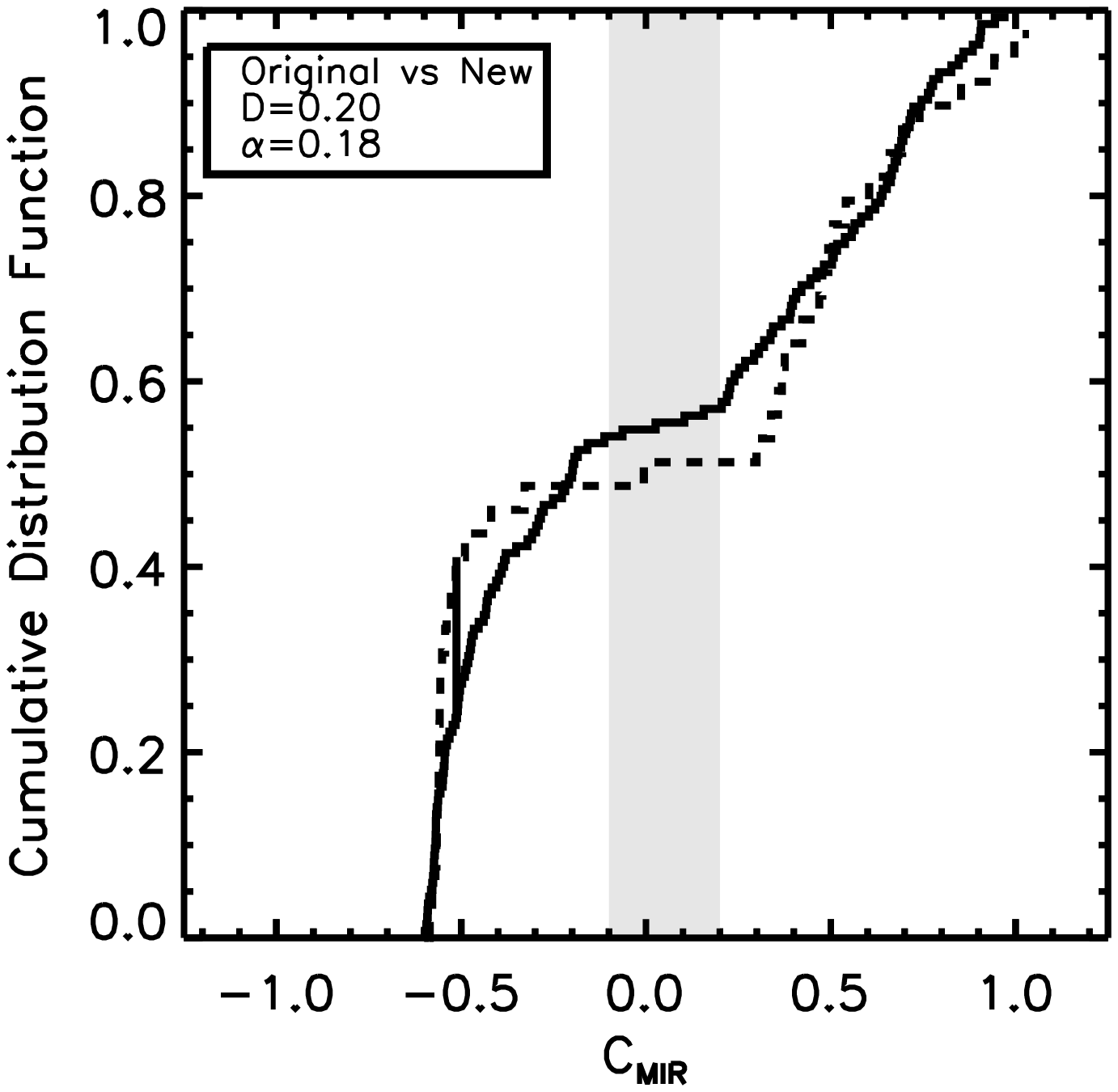}
  \caption{{\it Left}: The colorspace distribution of the compact group galaxies, with symbol color illustrating whether a galaxy is blueward (blue), in (green), or redward (red) of the gap. {\it Right}: KS test comparing the original sample of 12 HCGs (dashed line) with the new sample of 37 CGs (solid line). These two samples are very similar, with a fairly high probability of being drawn from the same parent distribution, 18\%.\label{idcompare}}
\end{figure}

The plots in Figure \ref{expks} show the results of the Kolmogorov-Smirnov (KS) test comparing the rotated and unwrapped color distributions with a model of a uniform distribution, where $D$ is the maximum difference between the sample and the model, and $\alpha$ gives the probability that the model matches the data. While we do not expect any particular sample to have a uniform distribution, using a model of a uniform distribution as a baseline facilitates comparisons between environments. The grey region indicates the canyon, which is clearly manifested in this plot as the flat portion of the cumulative distribution function (CDF). The canyon is less extended in this larger data set than the gap seen in the original sample, but it is still present, and $\alpha$ is small enough that we can conclusively state that the compact group sample is not drawn from a uniform distribution in colorspace. The two methods of unraveling the MIR colors yield the same result in the KS test. Further analysis uses the unwrapped colors, because the canyon galaxies as defined by the unwrapped sample are more cleanly defined in colorspace (shown in Figure \ref{idcompare}). Excluding the five RSCGs that are embedded in larger structures (21, 44, 66, 67, and 68) increases $\alpha$ by an order of magnitude, but does not change the result.

\subsection{Subsamples}
To investigate how the compact group environment affects galaxy evolution, we investigated the colorspace distribution of various subsamples. If a clear trend with physical properties were found, it would provide a clear picture of how environment affects galaxy evolution in compact groups.
\subsubsection{Original vs New}
To insure that the properties of the expanded sample are consistent with the original sample, we compare it with the original sample. The CDFs of the new and original sample are shown on the right in Figure \ref{idcompare}, which illustrates that the new sample is consistent with the original sample. The difference in the gap seen in the original sample and the canyon defined by the full sample (highlighted by the shaded region) is clearly illustrated by the different ranges of the flat portion of the CDF -- the canyon is narrower and slightly less pronounced than the original gap.

\subsubsection{HCGs vs RSCGs}
\begin{figure}
  \plottwo{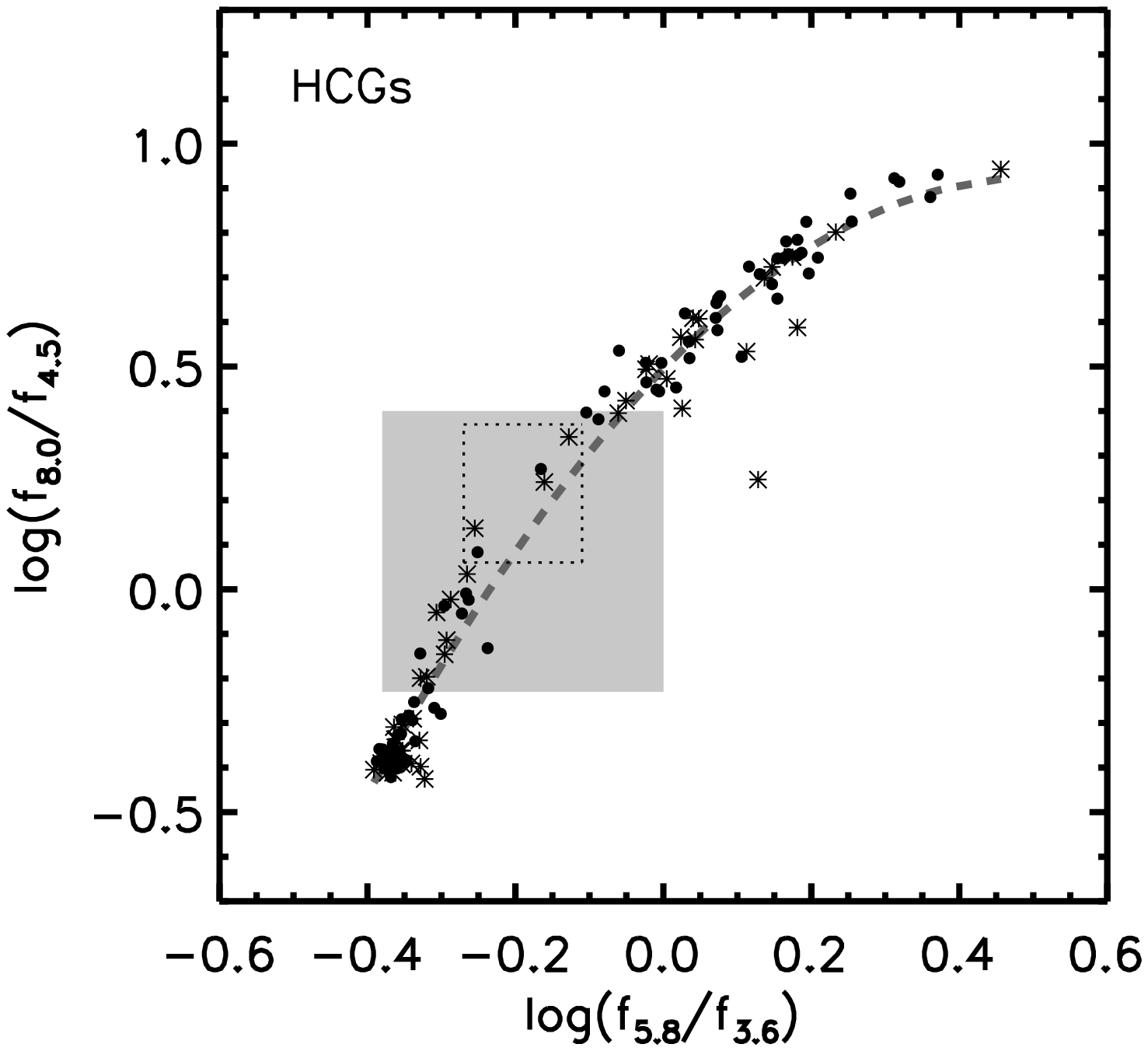}{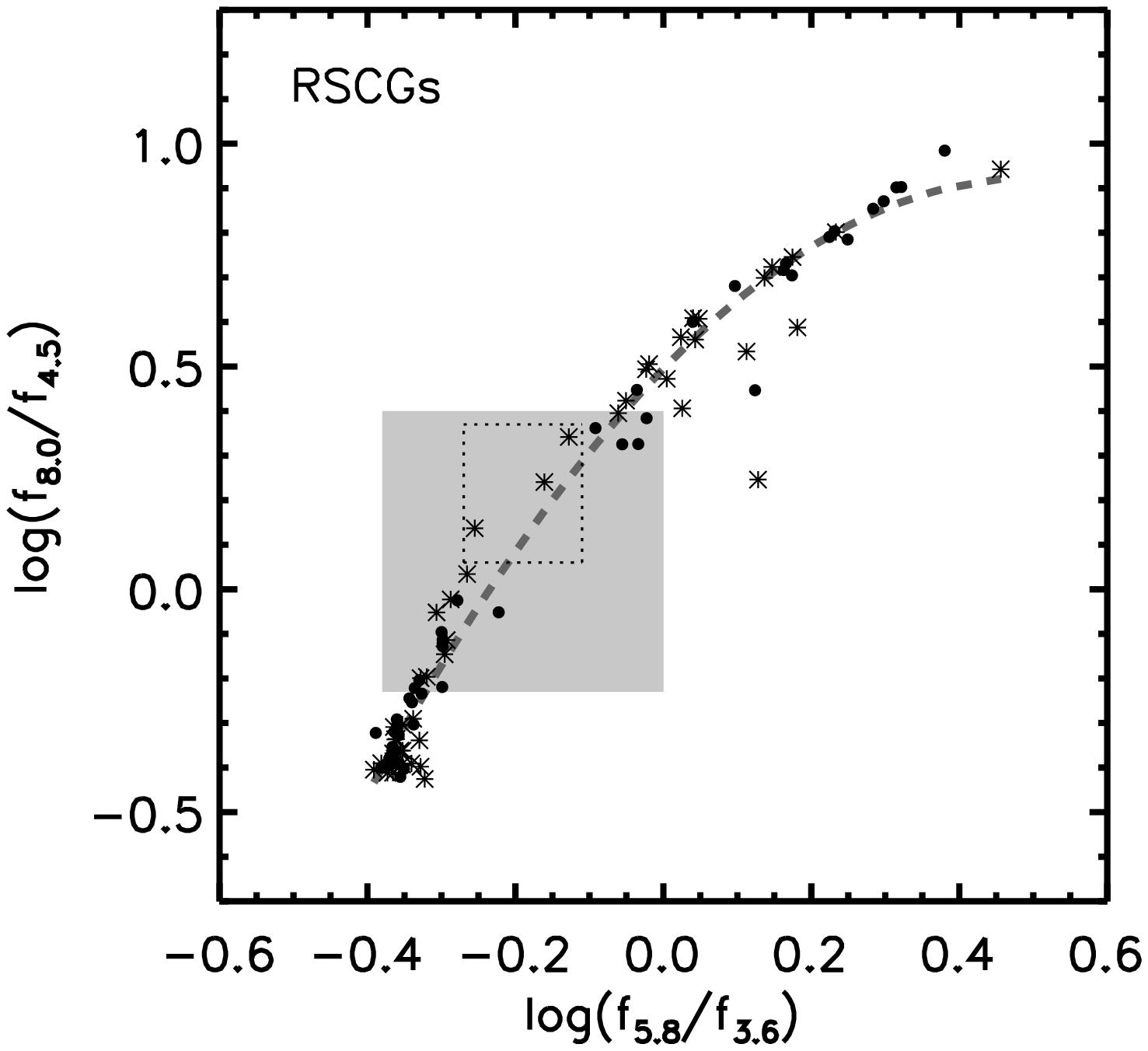}
  \caption{{\it Left}: The colorspace distribution of {\it left}: the 124 HCG galaxies and {\it right}: the 50 RSCG galaxies.\label{hcgsrscgs}}
\end{figure}

To determine whether the different selection criteria of the HCG and RSCG catalogs find groups with different properties, we compare groups from the two catalogs. The colorspace distribution of the HCG galaxies and RSCG galaxies are shown in Figure \ref{hcgsrscgs}. We see that the HCGs and RSCGs occupy colorspace very similarly.

\subsubsection{Separated by Physical Properties}
\begin{figure}
  \plottwo{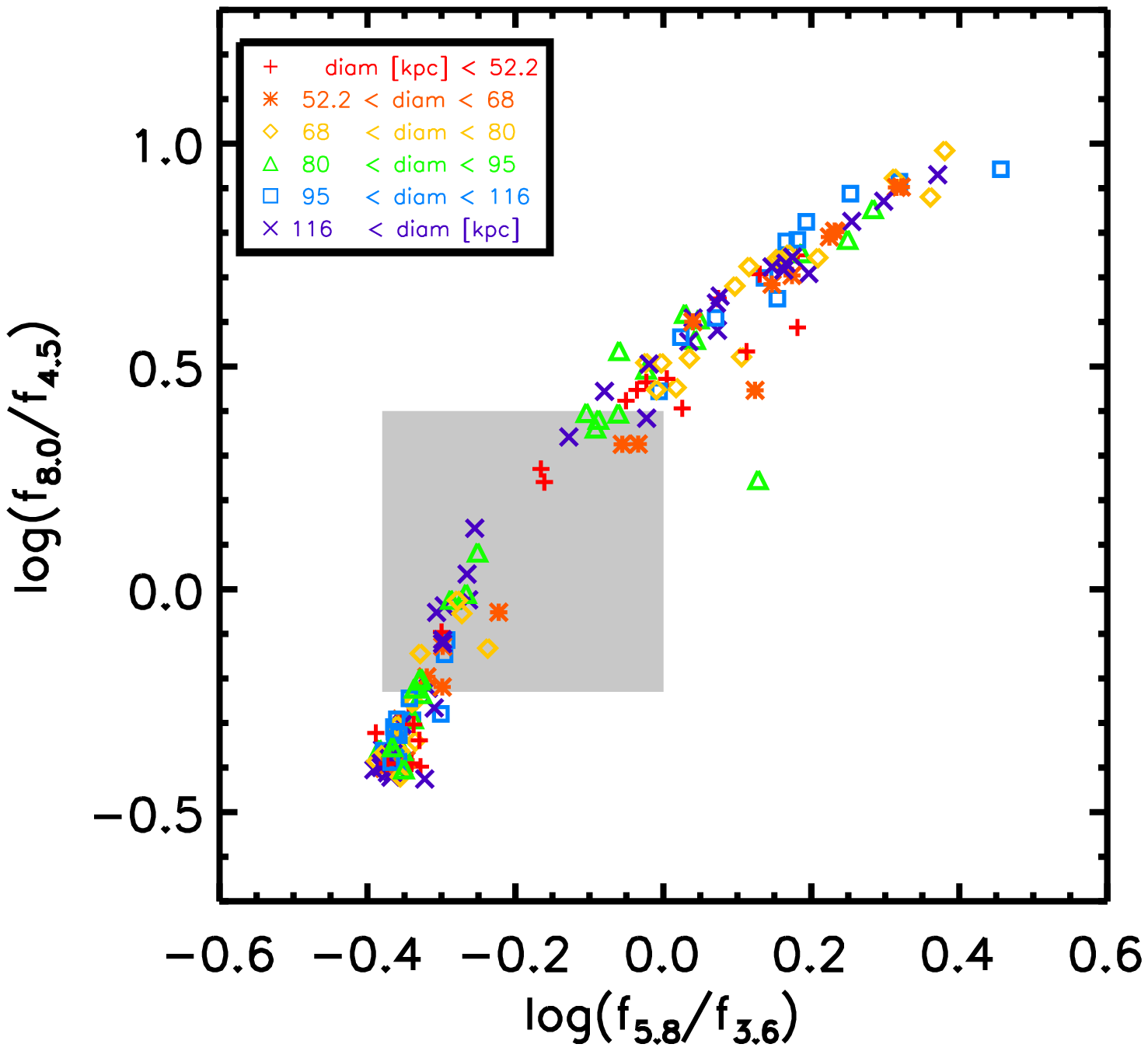}{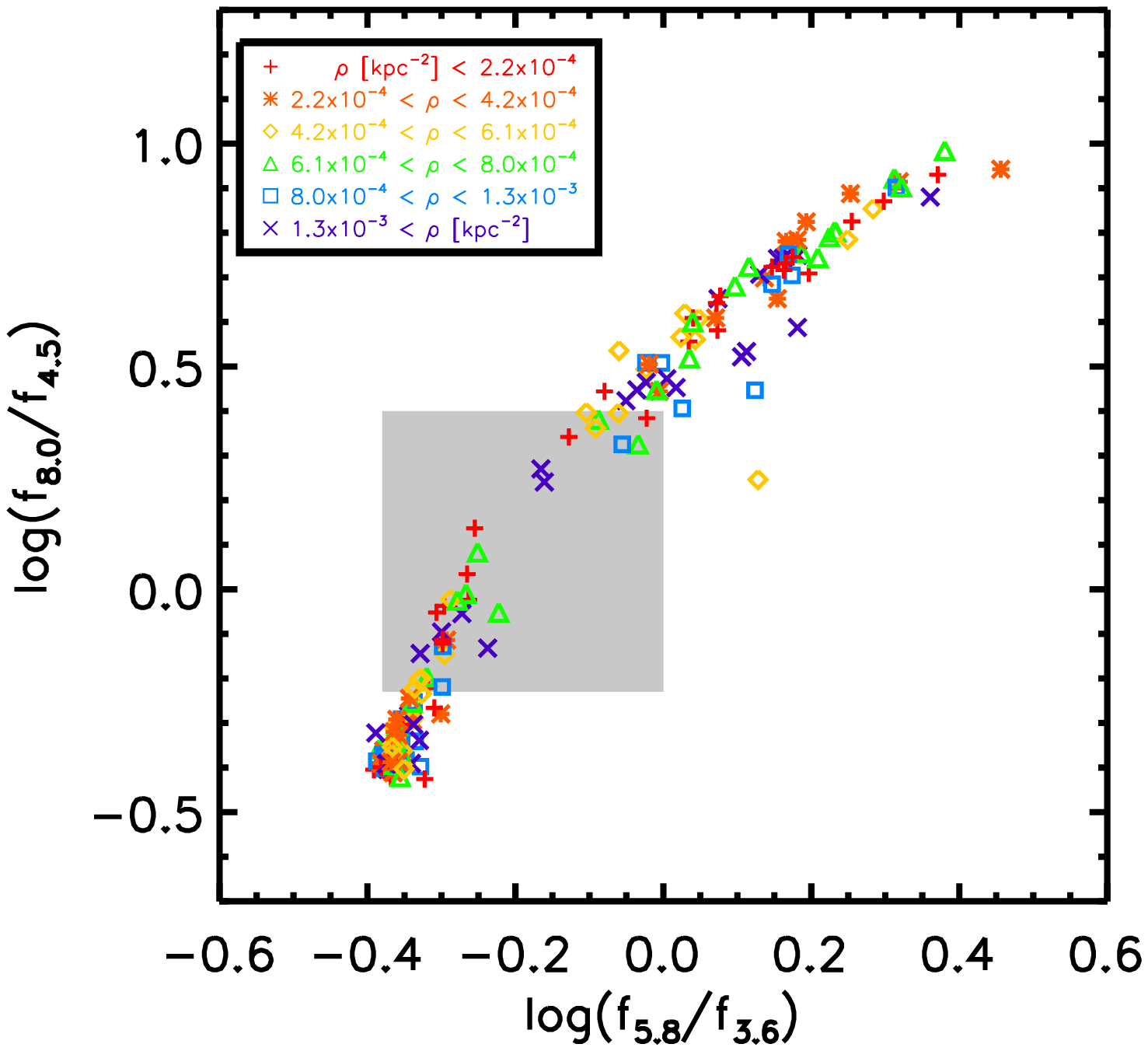}
  \caption{The colorspace distribution of the full sample, broken into subsamples based on {\it left}: projected physical diameter and {\it right}: projected physical density. Neither of these plots show a trend, indicating that these parameters do not play a key role in determining a galaxy's location in colorspace within the range represented by the compact group sample.\label{bydiam}}
\end{figure}

As compact group galaxies occupy colorspace differently than galaxies in other environments \citep[shown in Table \ref{ksfull} and \S\ref{compcol}, as well as][]{walker10}, we expect to see a trend with physical properties of compact groups, especially the properties which differentiate them from other environments. Thus we binned the groups by projected physical diameter and projected physical number density and examined colorspace as a function of these properties, as shown in Figure \ref{bydiam}. Both of these properties could be an indicator of how frequently or intensely interactions occur, which could plausibly affect the triggering of star formation and/or transformation from activity to quiesence. However, as these figures show, there does not seem to be any trend with these properties. Further investigation of the CDFs of the subsamples reveals that the shape of the subsample's CDF does not correlate with either of these properties. This will be discussed further in \S\ref{subsdiscuss}.

\subsection{Comparison Samples}\label{compcol}
\begin{figure}
  \plotone{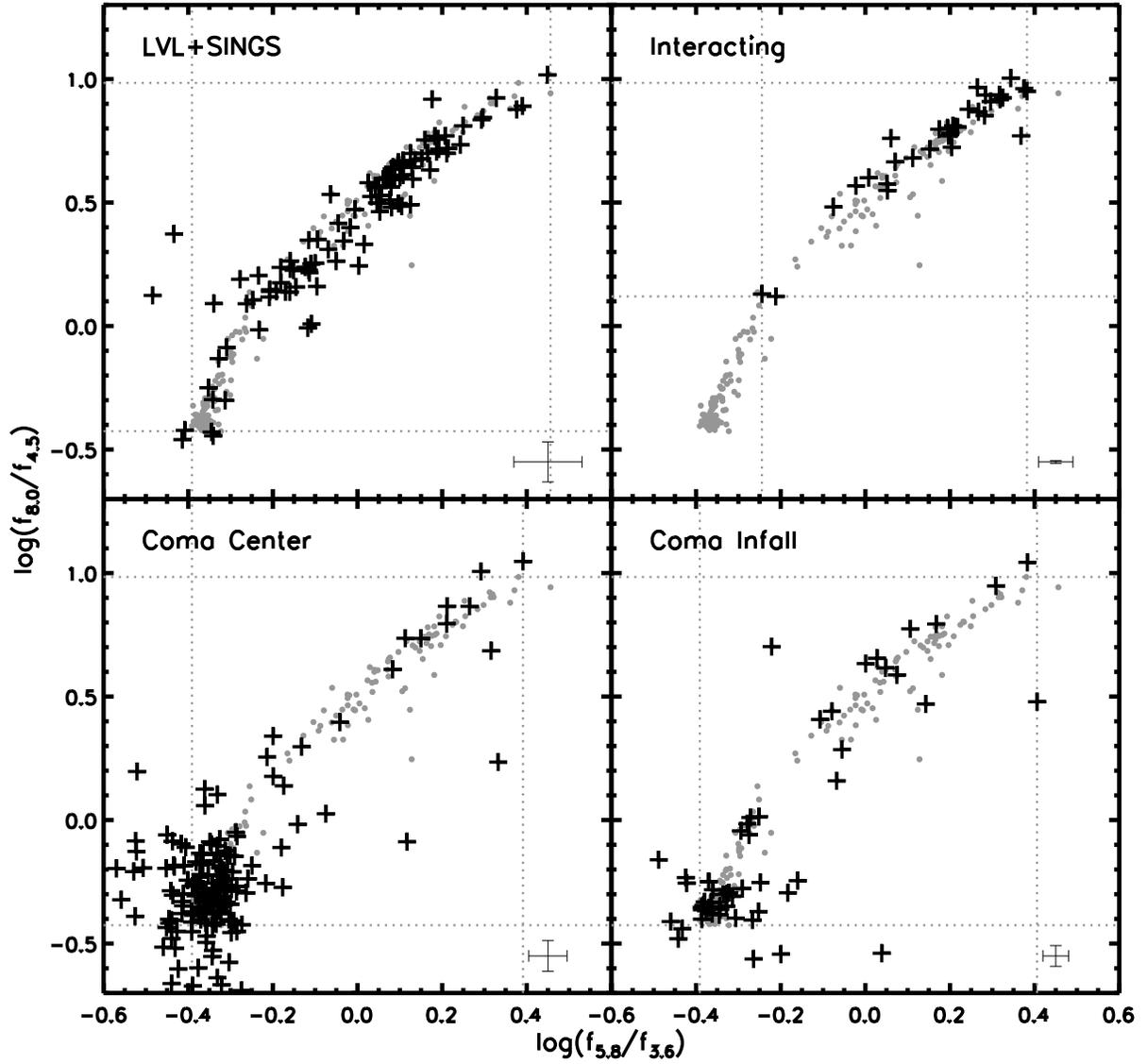}
  \caption{Colorspace distribution of the comparison samples (black plus signs) overlaid on the compact group sample (grey dots).\label{compcols}}
\end{figure}

The colorspace distributions of our comparison samples are shown in Figure \ref{compcols}. As this figure illustrates, the LVL+SINGS sample spans almost the exact same region of colorspace as the compact group sample, though there is no evidence for a canyon. The interacting sample does not occupy the blue region of colorspace, indicating that this sample does not contain any galaxies dominated by stellar emission. In contrast, galaxies in the center of the Coma cluster predominantly fall in the blue region of colorspace, with only a few galaxies whose colors indicate activity. Like the Coma center, the infall region of the Coma cluster shows a concentration of galaxies with normal stellar colors, but also contains galaxies with colors indicative of activity and exhibits an underdensity of points in the same region as the canyon in the compact group sample. We performed two-distribution KS tests comparing these samples with the compact group sample; the results are given in Table \ref{ksfull}.

It is important to consider the colorspace distributions of the samples in the context of the morphologies of their galaxies, given in Figure \ref{morphhist} for the compact group and field samples. As discussed in \citet{walker10}, untangling the effects of morphology on MIR color is non-trivial. To some extent, morphology and MIR color are expected to track each other; e.g., late-type spirals are typically starforming and therefore expected to have red MIR colors. Therefore, it could be the case that the different morphological types that dominate each of the comparison samples is driving the KS test results. For example, the interacting galaxy sample has no E/S0 galaxies by selection, and the morphology-density relation means that primarily quiescent E/S0s are found in the Coma Core sample. Unlike these two samples, both the compact group and LVL+SINGS samples span the range of morphological types. While MIR color tracks morphology quite well in the compact group sample, this is not the case for the LVL+SINGS galaxies \citep{walker10}. Thus, while the dearths seen in both colorspace and morphological distribution for the compact group sample may be caused by the same evolutionary process, the causal relationship is unclear.

\section{COLOR-MAGNITUDE DIAGRAMS}
\begin{figure}
  \plottwo{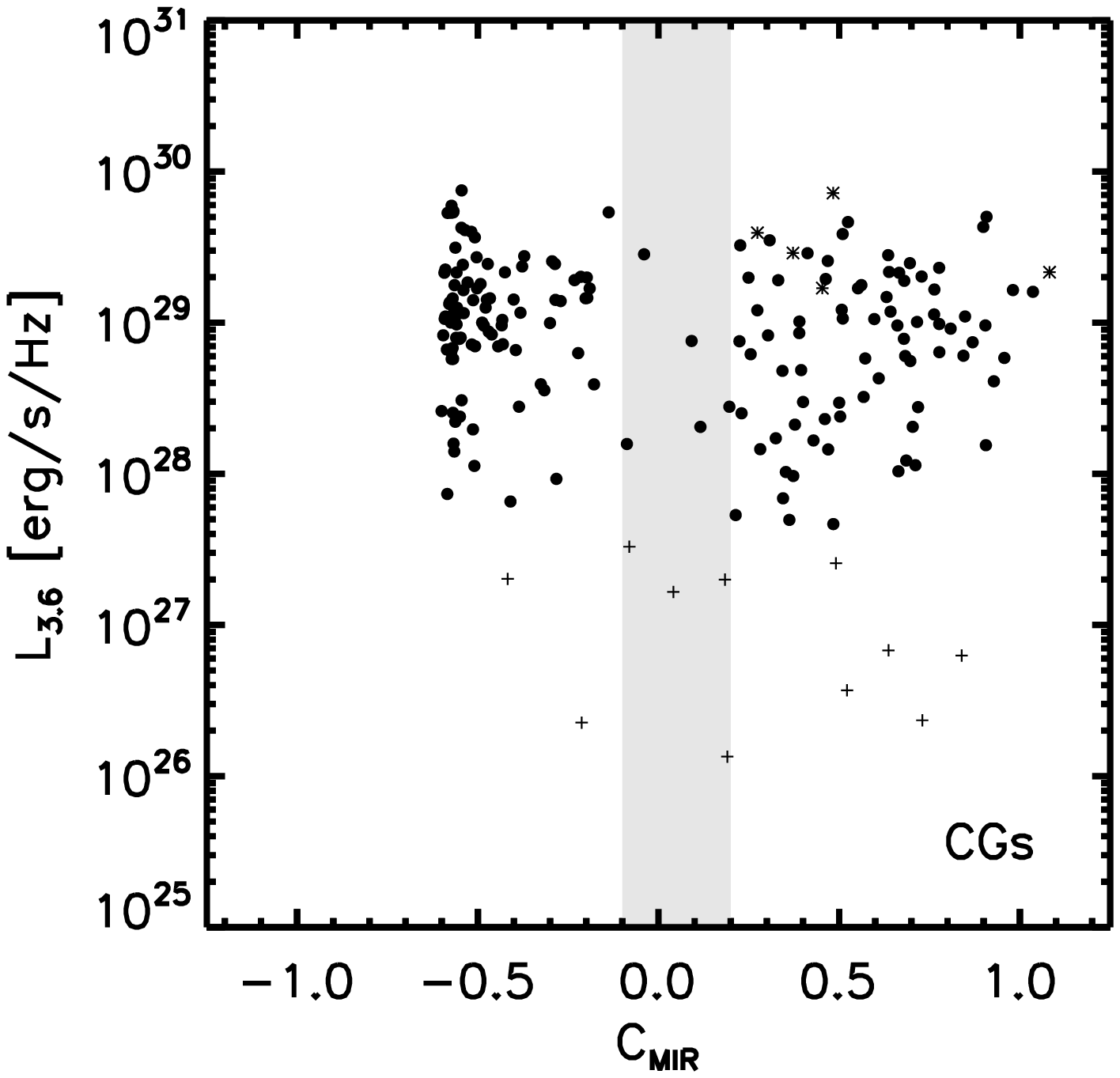}{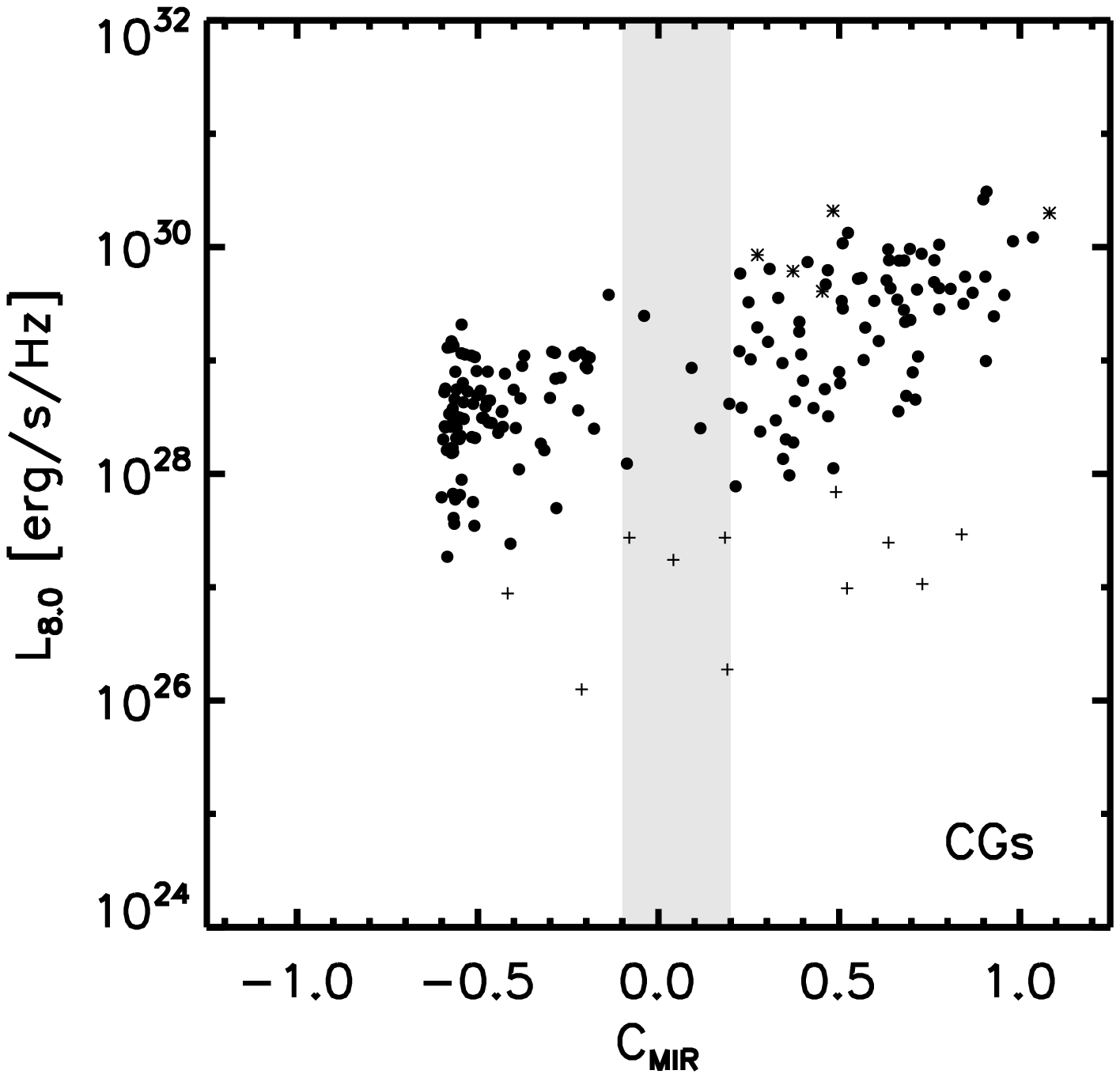}
\caption{$L_{3.6}$ ({\it left}) and $L_{8.0}$ ({\it right}) color-magnitude diagrams for compact group galaxies. The plus signs represent galaxies below the luminosity cut (as discussed in \S\ref{compsamples} and Figure \ref{lumhist}), the asterisks indicate saturated galaxies. The shaded region indicates the canyon in IRAC colorspace.\label{cgcmd}}
\end{figure}
\begin{figure}
  \plottwo{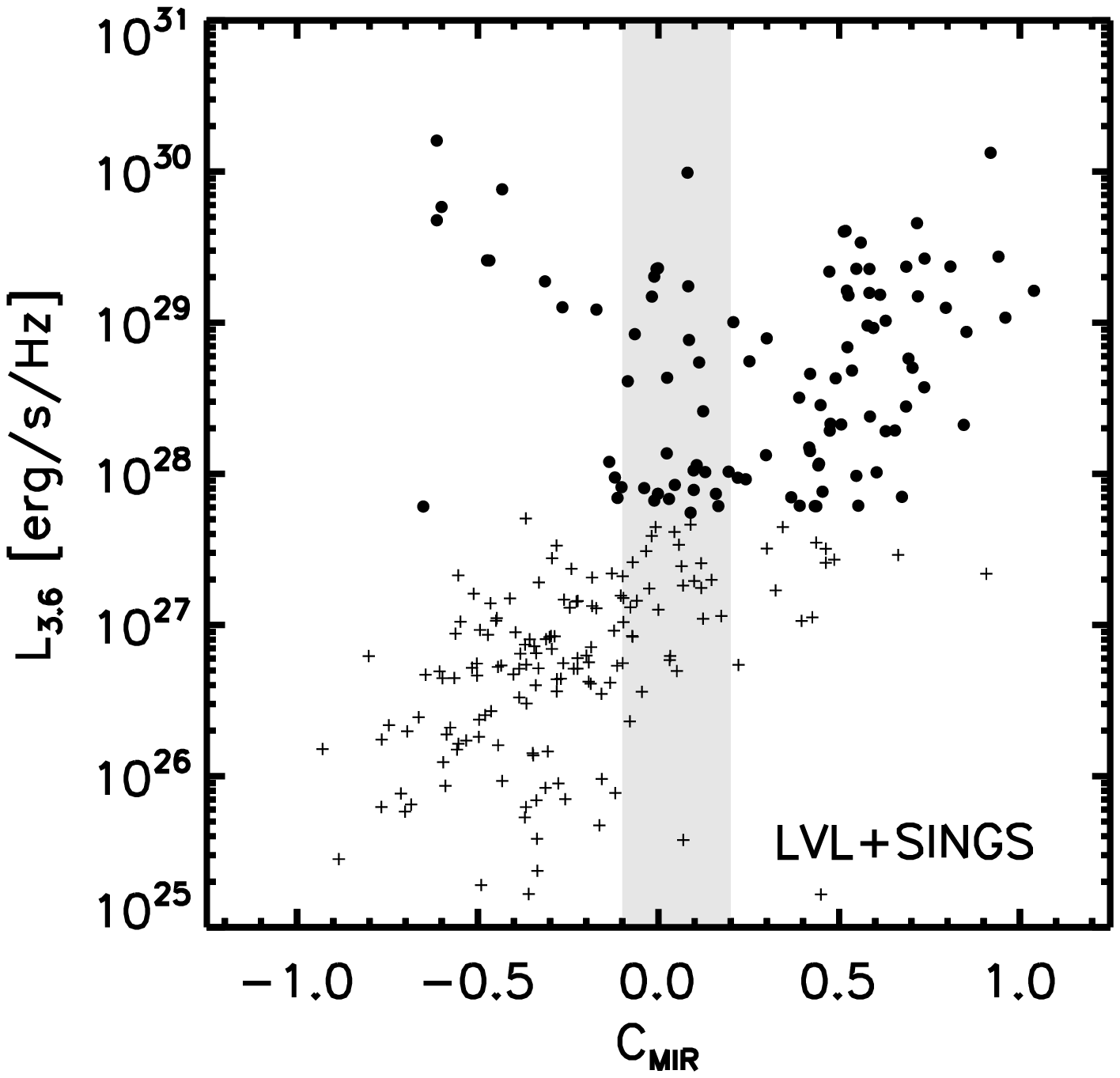}{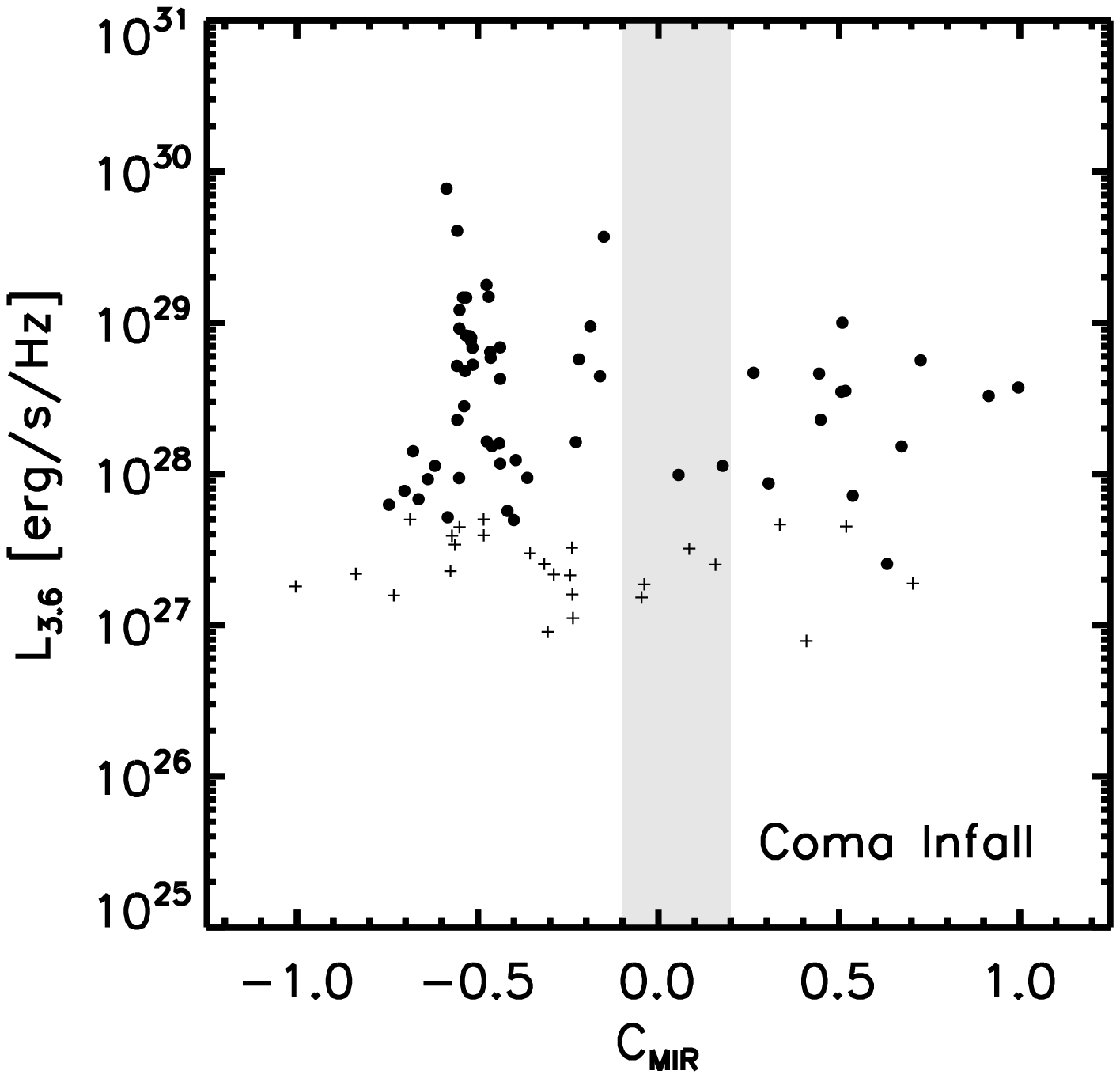}
\caption{$L_{3.6}$ color-magnitude diagrams of galaxies in {\it left}: the field sample of LVL+SINGS and {\it right}: the infall region of the Coma cluster. The plus signs represent galaxies below the luminosity cut (as discussed in \S\ref{compsamples} and Figure \ref{lumhist}). The shaded region indicates the canyon in IRAC colorspace.\label{3.6cmd}}
\end{figure}
\begin{figure}
  \plottwo{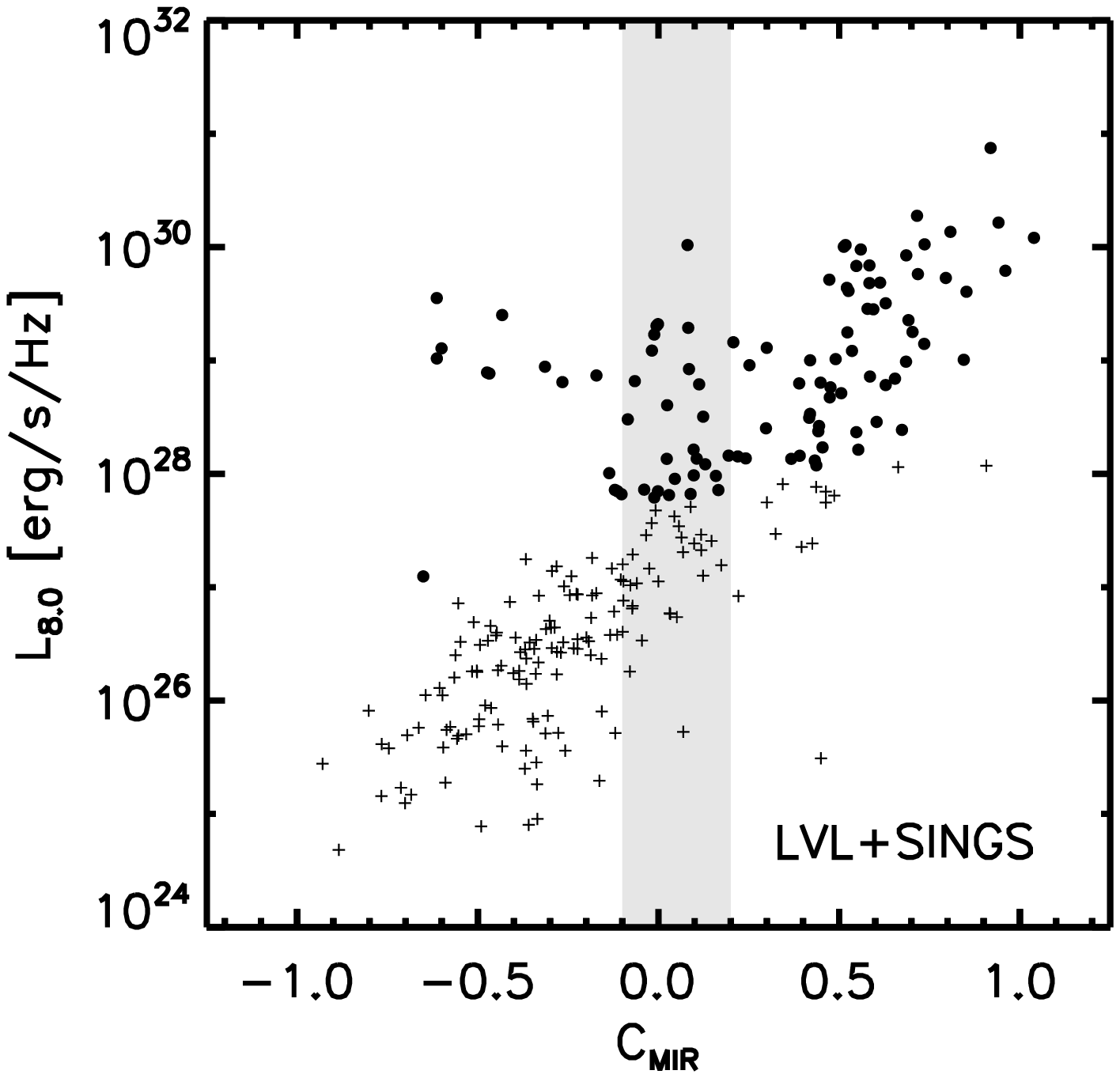}{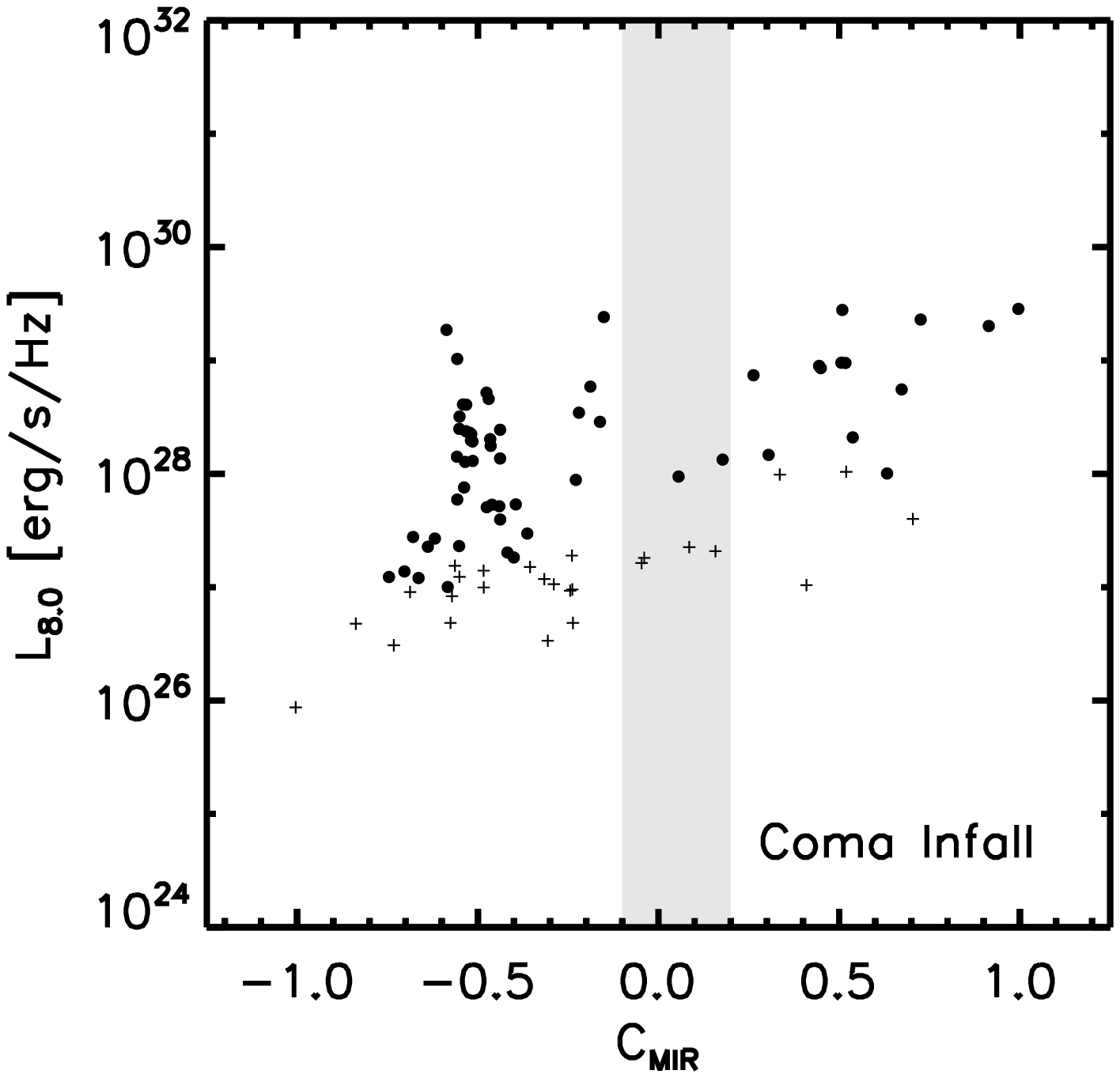}
\caption{$L_{8.0}$ color-magnitude diagrams of galaxies in {\it left}: the field sample of LVL+SINGS and {\it right}: the infall region of the Coma cluster. The plus signs represent galaxies below the luminosity cut (as discussed in \S\ref{compsamples} and Figure \ref{lumhist}). The shaded region indicates the canyon in IRAC colorspace.\label{8.0cmd}}
\end{figure}

Galaxy evolution is tied to galaxy size and buildup. If the compact group galaxies show a trend with mass or activity, this would be revealed in a color-magnitude diagram (CMD). Comparison of the CG CMD with LVL+SINGS and Coma Infall (shown in Figures \ref{cgcmd}, \ref{3.6cmd} and \ref{8.0cmd}) provide insight into galaxy properties. All three samples show a trend for MIR-red galaxies to be more luminous. This is not surprising, as we would expect that galaxies with activity would have more warm dust and PAH emission, thus increasing their $L_{8.0}$. One notable feature is the dearth of intermediate-luminosity, MIR-blue galaxies in the field sample, though this could be a selection effect, specifically that the SINGS sample was chosen to include ``interesting'' galaxies, thus mid-luminosity, MIR-blue galaxies may not have been included.

\section{SPECTRAL ENERGY DISTRIBUTIONS}
\begin{figure}
  \plottwo{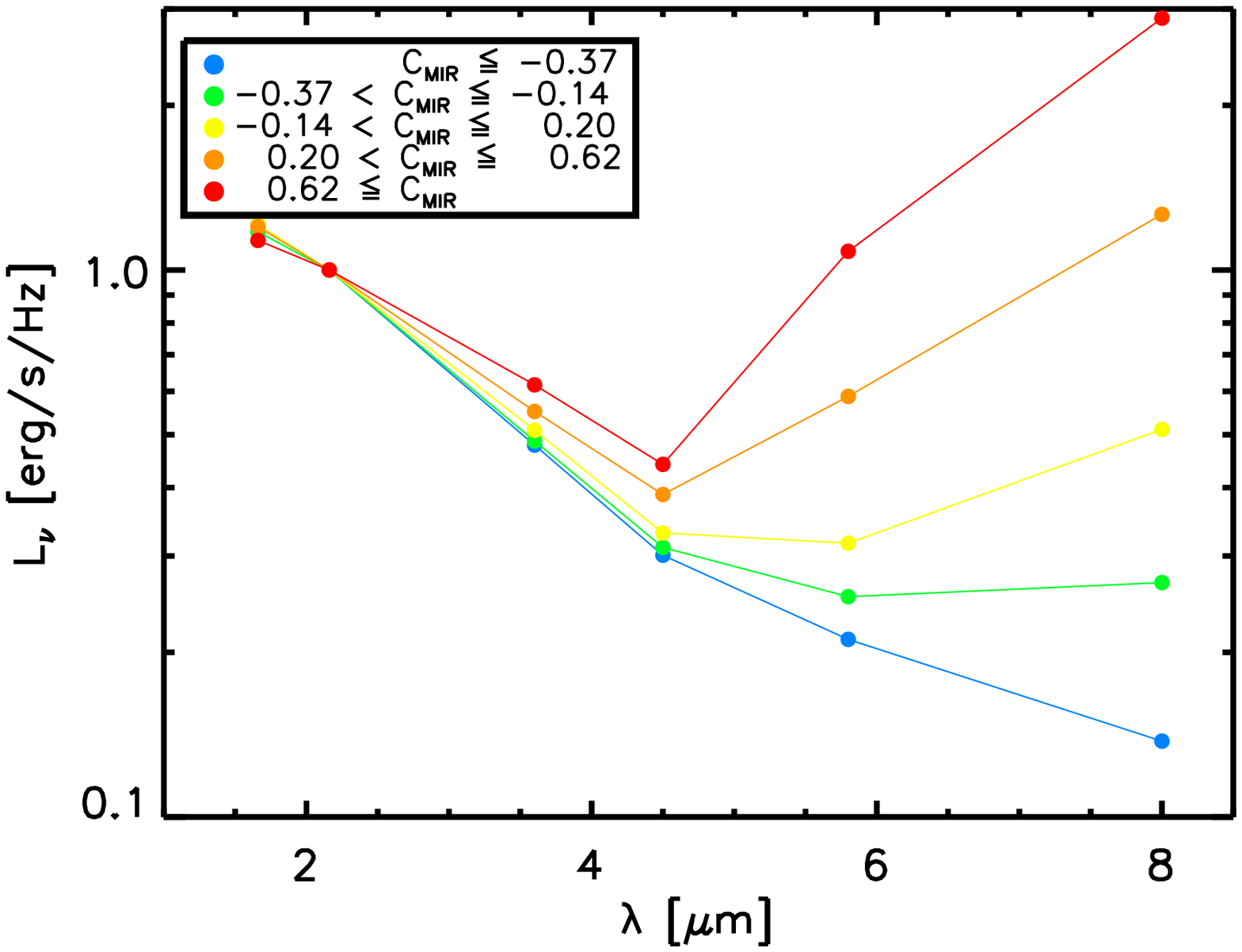}{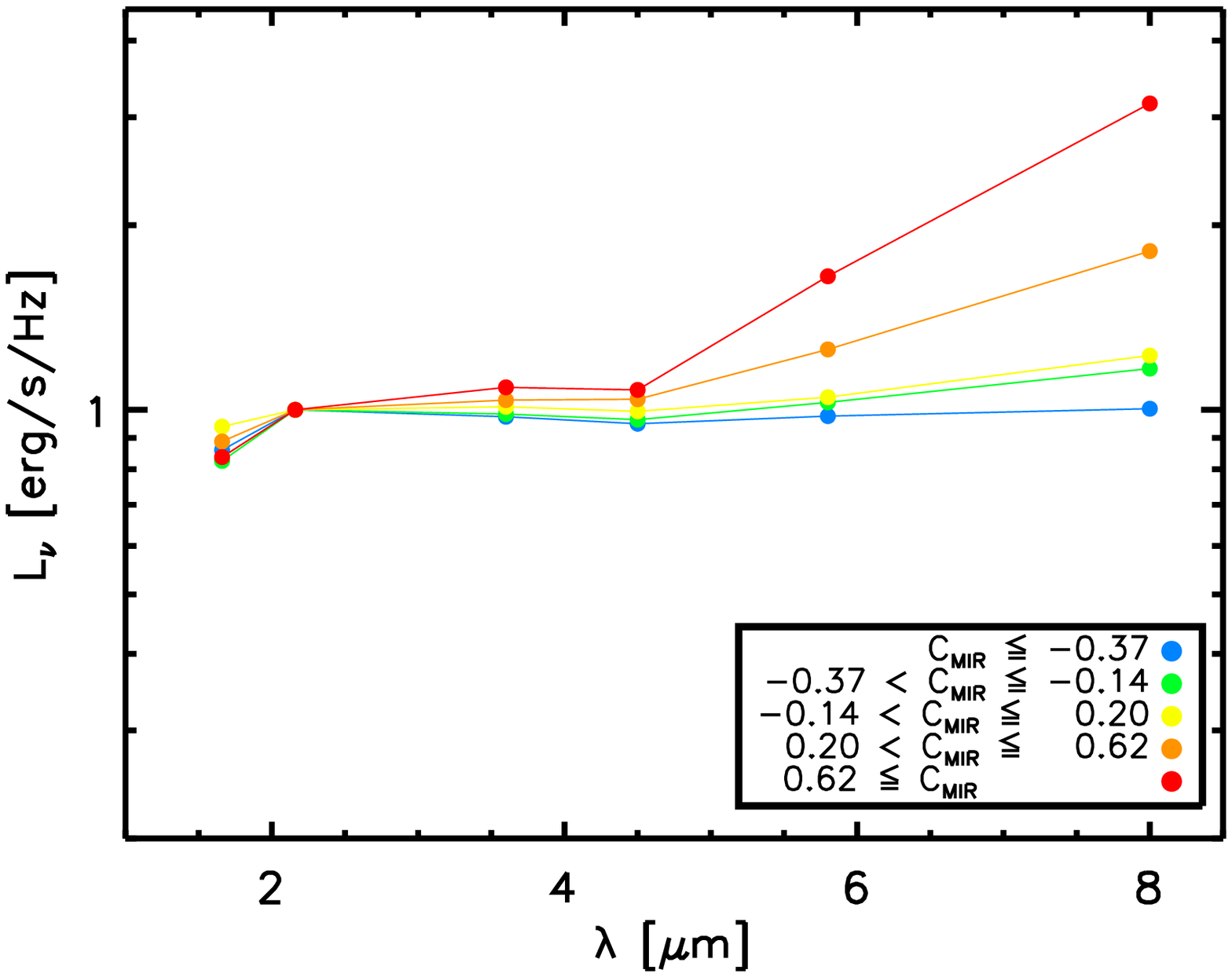}
\caption{Binned SEDs of compact group galaxies, normalized to K-band, as a function of location in MIR colorspace. The yellow SED is comprised of canyon galaxies. The plot on the left shows SEDs with contributions from both stars and dust. The plot on the right shows stellar-subtracted SEDs, allowing us to study the dust.\label{SEDs}}
\end{figure}

In order to investigate the relative contributions of amount and temperature of dust in compact group galaxies as a function of colorspace, we have binned SEDs of individual galaxies as a function of location in colorspace, shown on the left in Figure \ref{SEDs}. Since we are interested in the ISM, we extracted the contribution from stellar light by scaling a 5000K blackbody to the K-band emission. If the dust {\it properties} (e.g. temperature, composition, excitation) are the same in galaxies across colorspace, and the variation in MIR color is simply due to different {\it amounts} of dust and PAH emission, we expect the slope of the stellar-subtracted SEDs to be the same regardless of location in colorspace. If the different MIR colors are in fact dominated by varying dust properties, we expect the slope of the stellar-subtracted SEDs to change as a function of MIR color. As can be seen on the right in Figure \ref{SEDs} the stellar-subtracted SEDs do have clearly different slopes as a function of MIR color, which means that the range of MIR colors we see in the CG sample is not simply due to varying amounts of dust between galaxies. An analysis of the LVL+SINGS SEDs yields a similar result. Thus, the trend in IRAC colorspace can be attributed to the properties of dust in the galaxies, and the canyon reflects a range of moderate dust properties.

\section{DISCUSSION}
We have compiled a full sample of 49 compact groups comprised of 33 HCGs and 16 RSCGs. We see that the deficit (previously identified as a gap, see \S\ref{intro}) between galaxies with colors consistent with normal stellar emission (MIR blue) and galaxies with colors indicative of activity (MIR red) is persistent in the larger sample. We investigate physical trends as a function of colorspace and look at the structure of the colorspace distribution. We also compare the compact group colorspace distribution with samples of galaxies in other environments. We consider the SEDs of the galaxies to assess whether dust quantity or temperature changes as a function of colorspace.

\subsection{The Canyon}
The gap seen in the original sample of 12 HCGs \citep{johnson07,walker10} needs to be redefined in terms of the full sample presented here. While there is still an underdensity between galaxies whose colors are consistent with normal stellar populations and galaxies whose colors are indicative of activity, it is no longer the clear-cut gap seen originally. Now with better statistics, it is sparsely populated and thus more of a canyon. As shown in Figure \ref{colhist}, we quantified the location of the canyon via a histogram of the colorspace distribution. As the full sample has $\sim4\times$ more galaxies than the original sample, there is naturally a larger absolute number of galaxies in the canyon, though the relative number of galaxies are comparable (2.4\% of galaxies from the original sample, 2.8\% of galaxies from the expanded sample). The larger number of galaxies in this region will allow us to learn about properties of galaxies that exhibit these intermediate colors.

\subsection{Impact of Subsamples}\label{subsdiscuss}
The larger number of galaxies available in the full sample allows us to look for trends in colorspace. Comparing the HCGs with the RSCGs shows that they are consistent with being drawn from the same parent distribution. Breaking the full sample into subsamples based on projected physical diameter or density does not reveal any trends in colorspace. This is a puzzling result - we hypothesize that compact groups are quantitatively different from other samples in MIR colorspace due to our analysis of the compact group and comparison samples, yet we do not see any trends with these properties within the compact group sample. It appears that projected physical diameter and density are not the dominant properties in determining the colorspace location of compact group galaxies over the parameter ranges of this sample. It may simply be that the compact group environment is pre-selected to be compact and dense, and thus does not have a large enough variation in group diameters or densities to reveal trends in colorspace. However, \citet{johnson07} found that the original sample exhibits a trend in colorspace with group \ion{H}{1} richness, so this seems to be the crucial property in colorspace distribution. Regardless of a group's diameter or density, star formation relies on the presence of cold gas.

\subsection{Environment}
In agreement with previous work \citep{walker10}, the colorspace distribution of the compact group sample is statstically different from a control sample of ``field'' galaxies, a sample of interacting galaxies, and galaxies from the center of the Coma cluster. However, we cannot rule out the hypothesis that the compact group sample is drawn from the same parent distribution as the sample from the infall region of the Coma cluster. This supports the hypothesis that there is something special about this environment - one of high galaxy density where the neutral gas has not typically been fully preprocessed (meaning stripped, ionized, expelled, or otherwise modified through a galaxy interaction process). \citet{cluver11} found that galaxies in or near the canyon show anamolous $\mathrm{H_2}$ excitation, similar to that seen in the shock in Stephan's Quintet \citep[HCG92;]{cluver10}. This could be related to a rapid end to star formation in compact group galaxies, leading to the canyon observed in IRAC colorspace.

\subsection{SEDs}\label{sedsection}
The full sample revealed a curvature in colorspace not apparent in the original sample that was sparsely populated. The changing shape of the dust-subtracted SEDs as a function of MIR color indicate that the colors are not dominated solely by varying amounts of dust within the galaxies. Rather, the dust temperature distribution and/or PAH contributions change as a function of MIR color. We are unable to determine dust temperatures for these galaxies because the IRAC bands contain significant PAH features so we cannot disentangle the PAH emission from the dust temperature, and since the SEDs are not turning over, we do not cover the peak of the emission. We have looked at IRS spectra of compact group galaxies in different regions of IRAC colorspace, and found that the spectra of the bluest galaxies decrease with wavelength and show no PAH features. The PAH features then appear fairly quickly as you consider redder galaxies. They seem to be fully formed by the canyon region, and do not change significantly as you look redward. The primary change redward of the canyon seems to be the slope of the spectra (Walker et al. {\it in prep}). 
This suggests that galaxies undergoing activity have different dust compositions than quiescent galaxies.

\subsection{Conclusions}
We have determined that the distribution of compact group galaxies in MIR colorspace still shows an underdensity in the canyon region of colorspace. This distribution is most similar to that of the Coma infall region, implying a similarity in environment. The distribution of colors is caused by varying dust temperatures and PAH emission rather than varying amounts of dust. However, there are still many unexplained properties of this expanded sample.

\acknowledgments
K.E.J. gratefully acknowledges support for this paper provided by NSF through CAREER award 0548103 and the David and Lucile Packard Foundation through a Packard Fellowship. S.C.G. thanks the National Science and Engineering Research Council of Canada and The Ontario Early Researcher Award Program for support. J.C.C. was supported by the National Science Foundation through award 0908984. L.M.W thanks David Whelan for helpful discussions. We also thank the anonymous referee for their constructive comments. This research has made use of the NASA/IPAC Extragalactic Database (NED) which is operated by the Jet Propulsion Laboratory, California Institute of Technology, under contract with the National Aeronautics and Space Administration.

{\it Facilities:} \facility{Spitzer (IRAC)}.


\clearpage

\end{document}